\documentclass[12pt]{iopart}

\expandafter\let\csname equation*\endcsname\relax
\expandafter\let\csname endequation*\endcsname\relax

\usepackage{amsmath, xparse}
\usepackage{graphicx}% Include figure files
\usepackage{bm}% bold math
\usepackage{dcolumn}% Align table columns on decimal point
\usepackage[utf8]{inputenc}
\usepackage[T1]{fontenc}
\usepackage{mathptmx}
\usepackage{etoolbox}
\usepackage{xcolor}
\usepackage[colorlinks = true,
            linkcolor = blue,
            urlcolor  = blue,
            citecolor = blue,
            anchorcolor = blue]{hyperref}
\usepackage{cancel}
\usepackage{amssymb}
\usepackage{comment}

\usepackage{empheq}
\usepackage{soul}
\usepackage{pst-node}
\usepackage{tikz-cd}
\usepackage{relsize}
\usepackage{enumitem}

\newcommand{\indep}{\rotatebox[origin=c]{90}{$\models$}}
\begin{document}

\title{Neural network assisted electrostatic global gyrokinetic toroidal code using cylindrical coordinates}

\author{Jaya Kumar Alageshan$^1$, Joydeep Das$^1$, Tajinder Singh$^1$,
Sarveshwar Sharma$^{2,3}$, Animesh Kuley$^{1,*}$}

\address{$^{1}$Department of Physics, Indian Institute of Science, Bangalore 560012, India}
\address{$^{2}$Institute for Plasma Research, Bhat, Gandhinagar-382428, India}
\address{$^3$Homi Bhabha National Institute, Anushaktinagar, Mumbai, Maharashtra 400094, India}
\ead{akuley@iisc.ac.in}
\vspace{10pt}
\begin{indented}
\item[]August 2024
\end{indented}

\begin{abstract}

\noindent Gyrokinetic simulation codes are used to understand the microturbulence
in the linear and nonlinear regimes of the tokamak and stellarator core.
The codes that use flux coordinates to reduce computational complexities
introduced by the anisotropy due to the presence of confinement
magnetic fields encounter a mathematical singularity of the metric on the
magnetic separatrix surface. To overcome this constraint, we develop a
neural network-assisted Global Gyrokinetic Code using Cylindrical Coordinates
(G2C3) to study the
electrostatic microturbulence in realistic tokamak geometries. In particular,
G2C3 uses a cylindrical coordinate system for particle dynamics, which allows
particle motion in arbitrarily shaped flux surfaces, including the
magnetic separatrix of the tokamak. We use an efficient particle
locating hybrid scheme, which uses a neural network and iterative local
search algorithm, for the charge deposition and field interpolation. 
G2C3 uses the field lines estimated by numerical integration to 
train the neural network in universal function approximator mode to
speed up the subroutines related to gathering and scattering
operations of gyrokinetic simulation. Finally, as verification of the capability of
the new code, we present results from self-consistent simulations of linear
ion temperature gradient modes in the core region of the DIII-D tokamak.

\end{abstract}

%
% Uncomment for keywords
\vspace{2pc}
\noindent{\it Keywords}: Tokamak, PIC, Gyrokinetic, ITG, Neural Network, G2C3
%
% Uncomment for Submitted to journal title message
%\submitto{\NF}
%
% Uncomment if a separate title page is required
%\maketitle
% 
% For two-column output uncomment the next line and choose [10pt] rather than [12pt] in the \documentclass declaration
%\ioptwocol
%

%%%%%%%%%%%%%%%%%%%%%%%%%%%%%%%%%%%%%%%%%%%%%%%%%%%%%%%%%%%%%%%
\section{Introduction}
\label{Sec:Introduction}
%%%%%%%%%%%%%%%%%%%%%%%%%%%%%%%%%%%%%%%%%%%%%%%%%%%%%%%%%%%%%%%

\noindent Plasma turbulence in the scrape-off-layer (SOL) driven by microinstabilities
will play a crucial role in the plasma confinement and heat load to the tokamak wall.
Also, understanding the parasitic absorption of radio frequency waves in the SOL region
is still an open problem. It is believed that parametric decay instabilities will be a
plausible cause for such absorption \cite{Kuley09,Kuley10a,Kuley10b,Kuley15}.
Microinstabilities, like ion-temperature-gradient (ITG) and trapped electron modes (TEM)
are unstable due to the gradients in plasma temperature and density, and are known to
drive robust turbulence
activity\cite{Horton}. It is a critical challenge to capture both edge and core regions in an
integrated global simulation due to the shape complexity arising in the SOL region by
the divertor and X point. Presently, simulations within the SOL are primarily performed
with fluid and gyro-fluid codes based on Braginskii equations. Fluid codes, such as SOLPS
\cite{SOLPS} and BOUT++\cite{BOUT}, are widely used for SOL and divertor modelling. The significant
advantage of a fluid code is that it needs less computational effort than kinetic
approaches. Due to the plasma-wall interaction, the plasma is much colder in the
SOL region than the core and edge regions. Therefore, collisions play an essential role
in the SOL and influence turbulent transport. So, the fluid treatments based on
Braginskii's approach provides valuable insights into SOL turbulence. These codes keep
only a few moments and, therefore, cannot fully capture kinetic effects such as trapped
particles, nonlinear wave-particle interactions and suprathermal tail particles. Several
discrepancies have been reported between the experimental observations and fluid-based
simulations~\cite{Chankin,Canik}. Unlike fluid approaches,
the gyrokinetic simulation uses formulations that apply to a wide range of collisionality
regimes even though the collisional mean free path is not small compared to the parallel
scale length. Therefore, the fluid-based transport code will remain important
for the boundary plasma model. The global gyrokinetic simulation results are also
expected to provide a better understanding of boundary plasma simulation and validate the
experimental observations. Another promising code Gkeyll \cite{GkeyllES}, based on the
discontinuous Galerkin algorithm, has been recently applied to study the curvature-driven
turbulence in the open field line region and plasma wall interactions. Also, it is
extended to the nonlinear electromagnetic simulations in a helical open field line
system using National Spherical Torus Experiment (NSTX) parameters \cite{GkeyllEM}.\\ 

\noindent In the last three decades, our understanding of the microturbulence in the
tokamak core region has vastly improved, thanks to the development of several gyrokinetic
simulation codes such as GTC \cite{Tajinder24, Tajinder24a}, GYRO \cite{Waltz07}, 
C-GYRO \cite{CGYRO}, ORB5
\cite{Jolliet07}, GENE \cite{GENE}, etc. These codes use flux coordinates, which 
at the magnetic separatrix surface encounter a mathematical singularity in the metric.
To circumvent this problem, we have developed a new simulation code called G2C3 (Global
Gyrokinetic Code using Cylindrical Coordinates), similar in spirit to the XGC-1
\cite{Chang09}, GTC-X\cite{Sadhitro19} and TRIMEG \cite{Zhixin19}. Avoiding the flux
coordinate system allows G2C3, XGC-1, and TRIMEG to perform the gyrokinetic simulations
in arbitrarily shaped flux surfaces, including separatrix and X point in the tokamak. 
In G2C3, we have implemented fully kinetic (FK) and guiding center (GC) particle
dynamics, but XGC-1 and TRIMEG only have guiding center
particle dynamics. Also, TRIMEG uses a Fourier decomposition scheme, which is
computationally efficient for a single toroidal mode simulation. However, the toroidal
spectrum is broad for nonlinear simulations, and the simulation can be more expensive
than fully PIC code with field-aligned gather scatter operation. G2C3 uses field-aligned
gather-scatter operation to achieve field-aligned mesh efficiency. Recently, the GENE code
was also updated to GENE-X to incorporate the SOL region based on a flux coordinate
independent approach \cite{GENE-X}. The main focus for developing G2C3 is to couple the core and
SOL regions for understanding the electromagnetic turbulence using both the guiding
center and fully kinetic particle dynamics. \\

\noindent Neural network/machine learning methods have been employed in fusion research
to predict and control disruptions in discharges using
experimental data\cite{Tang,EPFL}, as a diagnostic tool to infer physical
quantities using data obtained from simulations~\cite{ProfileML}, to replace the
computationally expensive kinetic PIC simulations with reduced or surrogate
models~\cite{QLKNN}, and to model the collision operators~\cite{CollisionML1,CollisionML2}.
For the first time, G2C3 incorporates machine learning techniques within the global
PIC simulation. G2C3 uses a supervised multi-layered neural network to perform
interpolation along the magnetic field lines, with training data obtained via
numerical integration. This helps to perform scatter and gather operations.
Furthermore, G2C3 also uses a neural network to locate the particles within the mesh.
We benchmark our code by verifying the linear ITG mode in DIII-D core.\\

\noindent The work related to the ITG mode for the core region of DIII-D presented here
is electrostatic and provides only a first step towards achieving a desirable whole
plasma volume simulation and predictive capability. G2C3 reads the equilibrium fitting
(EFIT) \cite{LaoEFIT90} and IPREQ \cite{Deepti2020} data files generated from experimental
discharges. While the plasma ions are simulated with gyrokinetic marker particles,
the electron response is assumed to be adiabatic in this study. Other microturbulence modes
in the core and edge regions, such as TEM, ballooning modes,
pressure-driven magneto-hydrodynamic modes, and others, will be added in future work.  \\

\noindent In this article, we organize the material in the following form: after setting
the conventions and details of the equilibrium data, we briefly describe the
recipe to construct the simulation grids and the triangular mesh in
Sec.~\ref{Sec:Equilibrium}; the neural network-based gather-scatter module and
the triangle locator module are presented in Sec.~\ref{Sec:Interpolation}, with
error-analysis and convergence; the particle module in Sec.~\ref{Sec:Particledynamics}
describes the particle initialization and their governing dynamics; the finite
element-based gyrokinetic Poisson solver to estimate the electric field is
presented in Sec.~\ref{Sec:GKPoisson}. Finally, in Sec.~\ref{Sec:ITG} we describe how
the modules are used within a PIC cycle to simulate and analyze ion temperature
gradient-driven linear instability mode, with adiabatic electrons. The
conclusions are presented in Sec.~\ref{Sec:Conclusions}.

%%%%%%%%%%%%%%%%%%%%%%%%%%%%%%%%%%%%%%%%%%%%%%%%%%%%%%%%%%%%%%%
\section{Equilibrium magnetic field, coordinates and conventions}
\label{Sec:Equilibrium}
%%%%%%%%%%%%%%%%%%%%%%%%%%%%%%%%%%%%%%%%%%%%%%%%%%%%%%%%%%%%%%%

\begin{figure}[!ht]
    \centering        
        \includegraphics[scale=0.12]{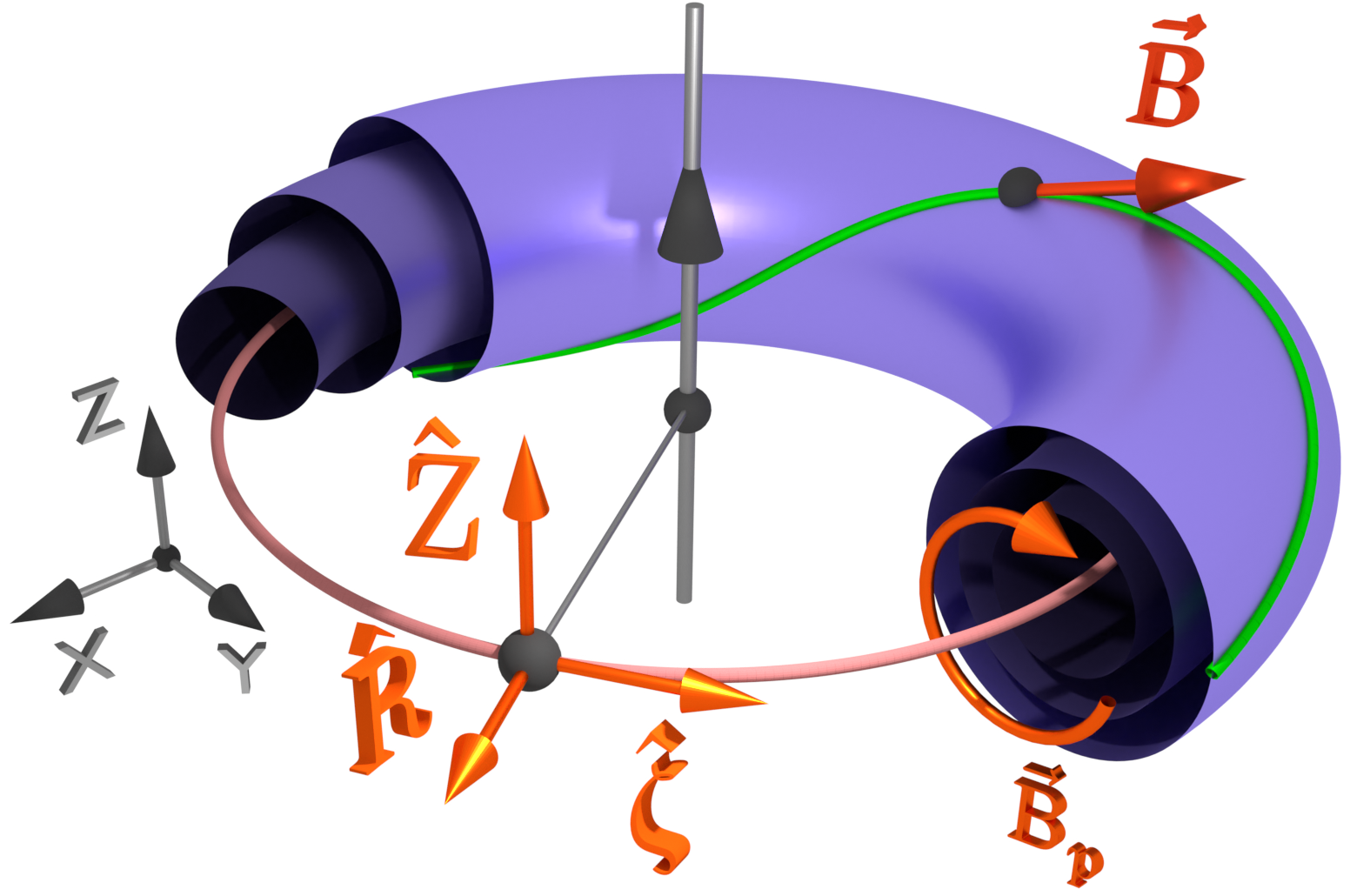}        
        \includegraphics[scale=0.2]{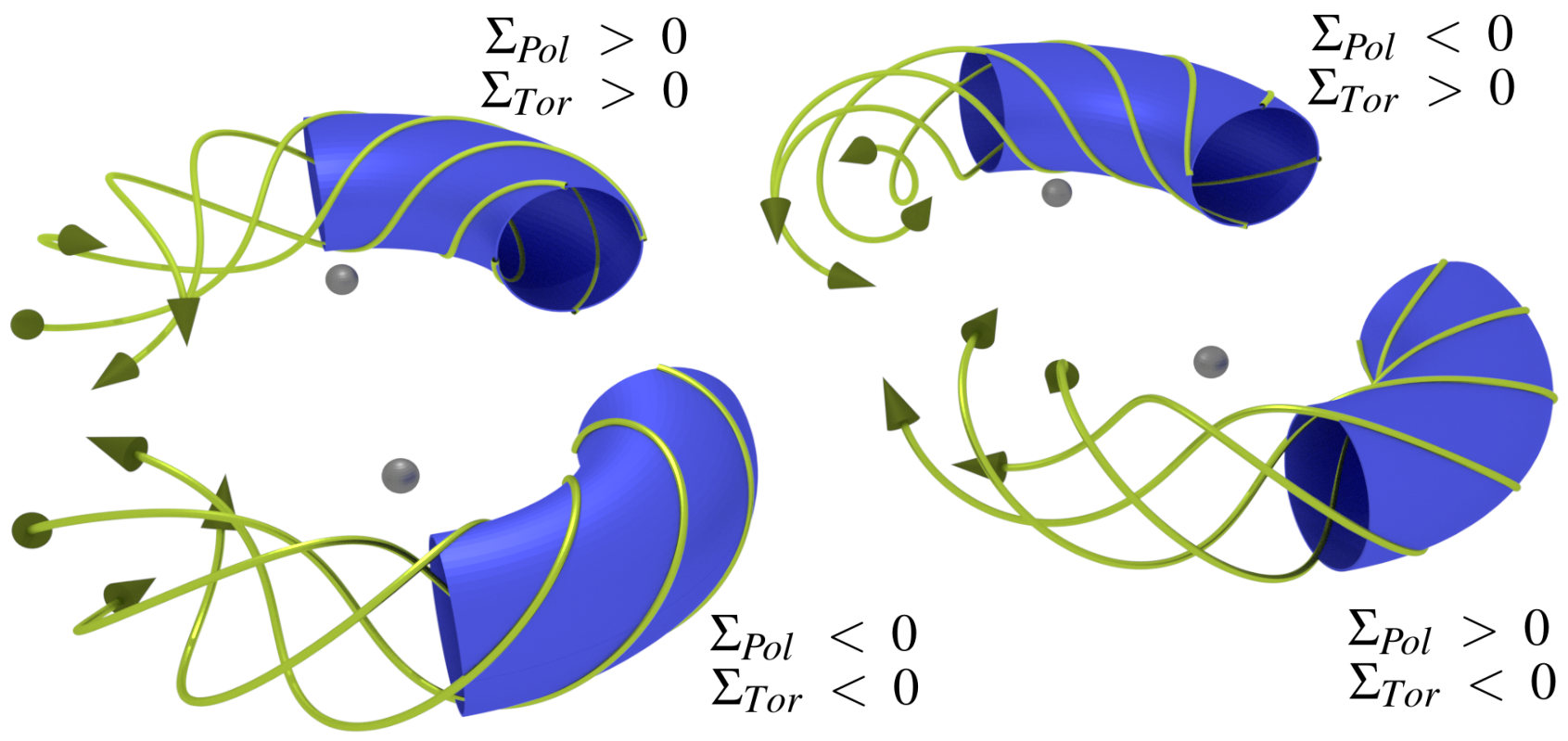}   
        {(a)} \hspace{6 cm}  {(b)}
    \caption{(a) Schematic of constant flux-function surfaces and a magnetic field line in different coordinate systems (Grey: Cartesian coordinate; Yellow: Cylindrical coordinate); (b) Configurations of magnetic field lines on a flux surface for different choices of $\Sigma_{Pol}$ and $\Sigma_{Tor}$.}
    \label{fig:coordinates}
\end{figure}

\noindent For an axisymmetric toroidal system, the poloidal flux function is symmetric in the toroidal direction, i.e., $\psi(R,\zeta,Z) = \psi(R,Z)$ such that 
$\psi$ is minimum at the magnetic axis, and ${\nabla}\cdot\vec{\mathbf{B}} = 0$ implies the equilibrium magnetic field can be written in an orthonormal coordinate system [see Fig.~\ref{fig:coordinates}] as:
\begin{eqnarray}
	\vec{\mathbf{B}} = \vec{\mathbf{B}}_{Pol} + \vec{\mathbf{B}}_{Tor} 
 %&=& \Sigma_{Pol} \; \left( \nabla\psi(R,Z) \times \nabla\zeta \right)
%	          + \Sigma_{Tor} \; \left( \frac{F_{Pol}(\psi)}{R} \; \hat{\mathbf{u}}^\zeta \right) \nonumber\\
	     = \Sigma_{Pol} \; \left( \nabla\psi(R,Z) \times \nabla\zeta \right)
	          + \Sigma_{Tor} \; \left( F_{Pol}(\psi) \; \vec{\mathbf{e}}^\zeta \right),
	          \label{eq:equilibrium}
\end{eqnarray}

%\begin{eqnarray}
%    \Sigma_{Pol}\; > \;0   \;\;\;  \Sigma_{Pol}\; > \;0  \;\;\;  \Sigma_{Pol}\; < \;0  \;\;\;  \Sigma_{Pol}\; < \;0  \nonumber \\
%    \Sigma_{Tor}\; > \;0   \;\;\;  \Sigma_{Tor}\; < \;0  \;\;\;  \Sigma_{Tor}\; > \;0  \;\;\;  \Sigma_{Tor}\; < \;0
%\end{eqnarray}

\noindent where $F_{Pol}(\psi)$ is the poloidal current function, such that $F_{Pol}(\psi)>0$; $\Sigma_{Pol} = sign\left( \vec{\mathbf{B}}_{Pol} \cdot 
(\nabla \psi \times \nabla\zeta) \right)$, 
$\Sigma_{Tor} = sign\left( \vec{\mathbf{B}}_{Tor} \cdot \vec{\mathbf{e}}^\zeta \right)$. Let the position vector in Euclidean space be
\begin{eqnarray}
   \vec{\mathbf{X}} = \{ x, y, z \} = \{R \: \cos\zeta, R \: \sin\zeta, Z \} \\
   \sigma^1 = R, \; \sigma^2 = \zeta, \sigma^3 = Z.
\end{eqnarray}
\noindent Therefore, the coordinate tangent vectors and the metric tensors are
\begin{eqnarray}
   \vec{\mathbf{e}}_i &=& \frac{\partial\vec{\mathbf{X}}}{\partial\sigma^i}, \;\;
   g_{ij} = \frac{\partial\vec{\mathbf{X}}}{\partial\sigma^i} \cdot 
	    \frac{\partial\vec{\mathbf{X}}}{\partial\sigma^j}
\end{eqnarray}

\noindent By defining contravariant basis vectors $\vec{\mathbf{e}}^R = \nabla R$, $\vec{\mathbf{e}}^\zeta = \nabla \zeta$, $\vec{\mathbf{e}}^Z = \nabla Z$, and $\vec{\mathbf{e}}^i = g^{ij} \; \vec{\mathbf{e}}_j$,  the metric component for this orthogonal basis, velocity, and magnetic field can be written as
\begin{eqnarray}
	g_{RR} = 1, \;\; g_{\zeta\zeta} = R^2, \;\; g_{ZZ} = 1, \;\;
	\text{and} \;\; g_{ij} = 0, \;\; \text{if} \; i\neq j
\end{eqnarray}
\begin{eqnarray}
	\vec{\mathbf{v}} \; = \; v^R \vec{\mathbf{e}}_R + v^\zeta \vec{\mathbf{e}}_\zeta 
	        + v^Z \vec{\mathbf{e}}_Z \; = \; v_R \vec{\mathbf{e}}^R 
	        + v_\zeta \vec{\mathbf{e}}^\zeta + v_Z \vec{\mathbf{e}}^Z
\end{eqnarray}
\begin{eqnarray}
	 \vec{\mathbf{B}} &=& \tilde{B}^R \; \vec{\mathbf{e}}_R 
	      + \tilde{B}^Z \; \vec{\mathbf{e}}_Z + \tilde{B}^\zeta \; 
	      \vec{\mathbf{e}}_\zeta 
	  = \tilde{B}_R \; \vec{\mathbf{e}}^R 
	      + \tilde{B}_Z \; \vec{\mathbf{e}}^Z + (R \; \tilde{B}_\zeta) \; 
	      \vec{\mathbf{e}}^\zeta
	      \label{eq:magneticbasis}
\end{eqnarray}
\noindent Equations ~(\ref{eq:equilibrium}) and ~(\ref{eq:magneticbasis}) provide the components of the magnetic field in cylindrical coordinates as 
\begin{eqnarray}
 \tilde{B}_R = -\frac{\Sigma_{Pol}}{\mathcal{J}} \; \frac{\partial \psi}
	     {\partial Z} ; \;\;
	\tilde{B}_Z = \frac{\Sigma_{Pol}}{\mathcal{J}} \; \frac{\partial \psi}
	     {\partial R} ; \;\; 
	\tilde{B}_\zeta = \Sigma_{Tor}\frac{F(\psi)}{R},
	\label{eq:magfiecomp}
\end{eqnarray}
\noindent where $\mathcal{J}$ is the Jacobian of the transformation, given by $\mathcal{J}=\sqrt{Det|g_{ij}|}$. \noindent The magnitude of the magnetic field is,
\begin{eqnarray}
	B = \sqrt{g^{ij} B_i B_j}= \sqrt{\tilde{B}^2_R + \tilde{B}^2_Z 
	      + \tilde{B}^2_\zeta}
\end{eqnarray}

\section*{Units and Normalization}
The basic units and normalization used in G2C3 are summarized as follows:
\begin{itemize}
    \item Mass: proton mass, $m_p$
    \item Charge: proton charge, e
    \item Magnetic field: on axis, $B_a$
    \item Length: Tokamak major radius, $R_0$
    \item Density: on axis electron density, $n_{e0}$
    \item Temperature: on axis electron temperature, $T_{e0}$
    \item Time: Inverse on axis cyclotron frequency of proton, $\omega_p^{-1}=m_pc/eB_a$
\end{itemize}

%%%%%%%%%%%%%%%%%%%%%%%%%%%%%%%%%%%%%%%%%%%%%%%%%%%%%%%%%%%%%%%
\subsection{Parameterizing the magnetic flux lines}
\label{Sec:Fluxlines}

\noindent In the presence of an external magnetic field, the plasma has a high degree of anisotropy, with vastly different length and time scales along the magnetic field lines and perpendicular to it. So to maintain the separation of scales and simplify calculations, it is convenient to define a coordinate system that encodes the field line geometry. We build the simulation grids which consider the anisotropy to enable efficient computation. The fields typically vary slowly in the $\parallel$-direction compared to the perpendicular directions. The magnetic field's global structure can be derived by integrating the differential equation of a curve which follows a magnetic field line. The local tangent to the magnetic field $d\mathbf{\mathcal{C}}_\mathbf{B}(s)/ds$, considering $\mathbf{\mathcal{C}}_\mathbf{B}(s)$ to be the trajectory along a magnetic field, can be written as:
\begin{eqnarray}
\frac{d\mathbf{\mathcal{C}}_\mathbf{B}(s)}{ds}\bigg|_{\text{along}\;\vec {\mathbf{B}}}=\frac{\vec{\mathbf{B}}}{B}=\hat b
\label{Eq:Fieldline_vector}
\end{eqnarray}

\begin{figure}[!ht]
	\centering{
	\includegraphics[scale=0.3]{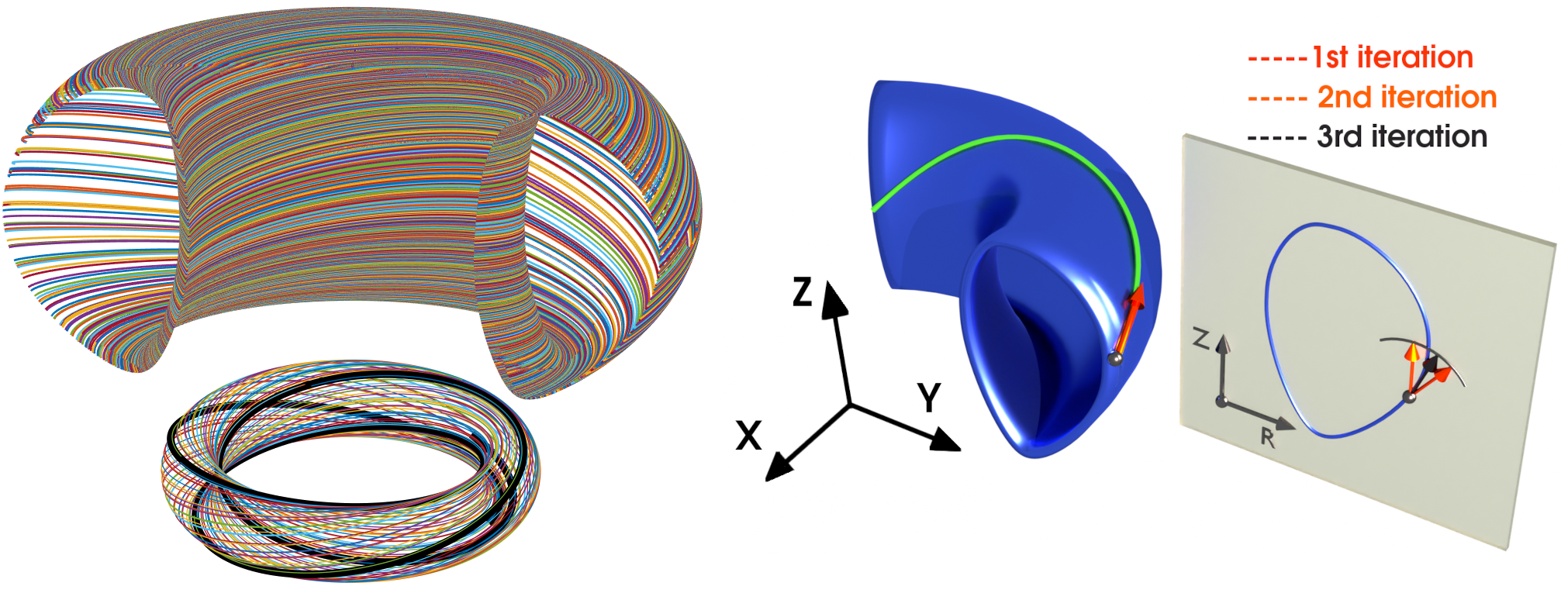} \\
    {(a)} \hspace{7cm} {(b)}}
	\caption{(a) A sectional (top) and complete (bottom) view of flux lines lying
            on the same flux surface, $\psi=0.28$;
	         (b) a schematic of how the numerically estimated field lines are constrained
          to lie on the flux surface and the iterative scheme we use to project the iteration
          points onto the flux surface, given by Eq.(\ref{Eq:UpdateCorrection}).}
    \label{Fig:FieldLines}
\end{figure}

\noindent We parametrize the flux line/curve using its natural length $s$, such that
$ds^2 = dx^2+dy^2+dz^2$. Then the flux-line is given by: $\mathbf{\mathcal{C}}_\mathbf{B}(s) = \left\{  x(s), \; y(s), \; z(s) \right\}$. If $\vec{\mathbf{B}}(\mathbf{X})$ is the magnetic field in Euclidean 3d space, i.e.
$\mathbf{X}\in \mathbb{R}^3$, then the magnetic flux line is the integral curve corresponding to $\hat b$
and this curve through a point $\mathbf{X}_0$ is given by:
\begin{eqnarray}
	\mathbf{X}(s) = \mathbf{X}_0 + \int_0^s \hat{b}\left(\mathbf{X}(\tilde{s})
	       \right) \; d\tilde{s}
	 \label{eq:Flux_Line}
\end{eqnarray}
\noindent The equation above for the flux line can be solved numerically by discretizing $s$ and using Euler's method as, 
\begin{eqnarray}
	\mathbf{X}(s_i) = \mathbf{X}(s_{i-1}) + \;
	    \Delta s \; \; \hat{b}\left(\mathbf{X}(s_{i-1})
	       \right) \nonumber
\end{eqnarray}
\noindent where $\Delta s = (s_i - s_{i-1})$. Figure~\ref{Fig:FieldLines}(a) shows the
numerically estimated trajectories for the DIII-D flux profiles using the Euler scheme
(and the $\psi$-invariance scheme described in Sec.~\ref{Sec:Psiinvariance}),
starting from all grid points of a given flux surface, $\psi=0.28$ at $\zeta=0$.
To numerically study the plasma in a tokamak,
the computational space is divided into a discrete set of poloidal planes,
specified by $\zeta_i = i\; \Delta\zeta$, where $i\in\{1,2,... N_p\}$, 
$\Delta\zeta = 2\pi/N_p$, and $N_p$ is the number of poloidal planes.
Any computation on points in between these poloidal planes, the point is
projected onto the nearest poloidal planes along the flux line given by
Eq.~(\ref{eq:Flux_Line}). But the estimation of $s$ on the $\zeta_i$'s is
non-trivial. If $d\zeta/ds \neq 0$ at all points along the flux line,
we can simplify the estimation of projected points by parametrizing
the curve using $\zeta$. Therefore,
\begin{eqnarray}
	\frac{d\mathbf{X}}{d\zeta} = \frac{ds}{d\zeta} \;
	    \frac{d\mathbf{X}}{ds} = \frac{ds}{d\zeta} \; \hat{b}, 
	    \label{eq:xszeta}
\end{eqnarray}
\noindent so the Euler scheme based linear projection operator that projects coordinates
onto $\zeta_i$ poloidal plane, $\mathcal{P}_{\zeta_i}$ is given by
\begin{eqnarray}
	\mathcal{P}_{\zeta_i} \mathbf{X} &=& \mathbf{X} + \int_\zeta^{\zeta_i} \;
	     \frac{d\mathbf{X}}{d\zeta} \; d\zeta, \nonumber\\
	     &\approx& \mathbf{X} + (\zeta_i - \zeta) \;
	     \left. \frac{d\mathbf{X}}{d\zeta}\right|_\zeta \;\; 
	     \label{eq:euclideanprojection}
\end{eqnarray}
\noindent Also,
\begin{eqnarray}
   \hat{b} = \frac{d\mathbf{X}}{ds} = \{ \; B_x/B \; , \; B_y/B \; , \; B_z/B \; \} \nonumber \\
        = \{ \; \frac{dR}{ds} \; \cos\zeta - R 
        \sin\zeta \; \frac{d\zeta}{ds} \; , %\nonumber \\
        \; \frac{dR}{ds} \; \sin\zeta + R \cos\zeta \; \frac{d\zeta}{ds} \; , %\nonumber \\
        \; \frac{dZ}{ds} \; \}      
        \label{eq:UnitProjection}
\end{eqnarray}
\noindent Comparing the first two components we get,
\begin{empheq}[right=\empheqrbrace]{align}
     \frac{d\zeta}{ds}& = (B_y \; \cos\zeta - B_x \; \sin\zeta )/(RB) \;
    = \; (\hat{\mathbf{e}}_\zeta . \vec{\mathbf{B}}) / (RB) \; = \; 
    \frac{\tilde{B}_\zeta}{RB},  \;\; \nonumber \\
    \frac{dR}{ds}& = (B_x \; \cos\zeta + B_y \; \sin\zeta )/B \;
    = \; (\hat{\mathbf{e}}_R . \vec{\mathbf{B}}) / B \; = \; \frac{\tilde{B}_R}{B}, \nonumber \\
    \frac{dZ}{ds}& = \frac{\tilde{B}_Z}{B}.
    \label{Eq:FieldLines}
\end{empheq}

\noindent Assuming $d\zeta/ds \ne 0$ (i.e. magnetic field is never purely poloidal), we can write $ds/d\zeta = (d\zeta/ds)^{-1}$ and 
$\Delta s \approx (R \; B / B_\zeta) \Delta\zeta$. Therefore,
\begin{empheq}[right=\empheqrbrace]{align}
	R \; \rightarrow& \;\; R + \Delta s \; \frac{dR}{ds} = R + \Delta\zeta 
	      \; \frac{R \; B_R}{\tilde{B}_\zeta} , \;\; \nonumber \\
	Z \; \rightarrow& \;\; Z + \Delta s \; \frac{dZ}{ds} = Z + \Delta\zeta 
	      \; \frac{R \; B_Z}{\tilde{B}_\zeta} , \nonumber \\
	\zeta \; \rightarrow& \;\; \zeta + \Delta s \; \frac{d\zeta}{ds} 
        = \zeta + \Delta \zeta,\nonumber \\
	s \; \rightarrow& \;\; s + \Delta\zeta \; \frac{R \; B}{\tilde{B}_\zeta}.
	\label{Eq:FieldLineUpdate}
\end{empheq}

%%%%%%%%%%%%%%%%%%%%%%%%%%%%%%%%%%%%%%%%%%%%%%%%%%%%%%%%%%%%%%%
\subsection{Invariance of \texorpdfstring{$\psi$}{TEXT} along the field lines}
\label{Sec:Psiinvariance}

\noindent As indicated in the schematic of the flux-surface in Fig.~\ref{Fig:FieldLines}(a), the field line should lie on the same flux surface, {\it i.e.} $\nabla_\parallel \psi = 0$. Conversely, the flux-surface constrains the magnetic field lines. So we need to ensure that the newly estimated point on the discrete flux-line has the same $\psi$ value. We impose this constraint at each update step in Eq.~(\ref{Eq:FieldLineUpdate}) by an iterative projection
onto the flux-surface as shown in Fig.~\ref{Fig:FieldLines}(b). We modify the update equation such that if
\begin{eqnarray}
\alpha = \arctan( \Delta Z / \Delta R ) \;, \;\; \text{and} \;\; 
r = \sqrt{\Delta R^2 + \Delta Z^2},
\end{eqnarray}
then
\begin{empheq}[right=\empheqrbrace]{align}
    \Delta R \; \rightarrow & \;\; r \; \cos(\alpha), \nonumber \\
    \Delta Z \; \rightarrow & \;\; r \; \sin(\alpha), \nonumber \\
    \Delta \psi \; \rightarrow & \;\; \left[ \psi(R+\Delta R,Z+\Delta Z) 
              - \psi(R,Z) \right], \nonumber \\
    \Delta \alpha \; \rightarrow & \;\; sign(\Delta \psi) 
            \; \frac{|\Delta \alpha|}{2},  \;\nonumber \\
        \alpha \; \rightarrow & \;\; \alpha + \Delta \alpha.
    \label{Eq:UpdateCorrection}
\end{empheq}
\noindent The above iteration is performed with initial $\Delta \alpha = \pi/2$, until $|\Delta\psi|$ is
less than a predefined cut-off value. This ensures that the field lines do not deviate by more
than the cut-off from the starting $\psi$ at all iterations.

%\noindent Furthermore, we can also implement the $\nabla_\parallel \psi = 0$ constraint using a gradient descent method, such that if 
%$\mathcal{E}:=\left[ \psi(R',Z') - \psi(R,Z) \right]^2$, then
%\begin{eqnarray}
%   R' \rightarrow R' - \eta_{GD} \; \frac{\partial \mathcal{E}}{\partial R'} \; = \; 
%         R' - 2 \eta_{GD} \; \left[\psi(R',Z')-\psi(R,Z)\right] \frac{\partial \psi(R',Z')}{\partial R'},  \nonumber \\
%   Z' \rightarrow Z' - \eta_{GD} \; \frac{\partial \mathcal{E}}{\partial Z'} \; = \; 
%         Z' - 2 \eta_{GD} \; \left[\psi(R',Z')-\psi(R,Z)\right] \frac{\partial \psi(R',Z')}{\partial Z'},
%\end{eqnarray}
%where $\eta_{GD} \in \mathbb{R}^+$ is a small {\it learning-parameter}. 
%But the gradient-based approach fails near the X-point, as $\nabla_\perp \psi \approx 0$.

%%%%%%%%%%%%%%%%%%%%%%%%%%%%%%%%%%%%%%%%%%%%%%%%%%%%%%%%%%%%%%%
\subsection{Grids for the core region}
\label{Sec:Grids}

\noindent We normalize and redefine $\psi$ generated using IPREQ \cite{Deepti2020} and EFIT \cite{LaoEFIT90} such that $\psi(R,Z)\geq0$ and $\psi(R_0,Z_0)=0$, where $(R_0,Z_0)$ is the position of the magnetic axis. Let $\psi_\times$ be the $\psi$-value along the flux surface through the X-point. Then we consider the annular region with $0 < \psi_1(R,Z) < \psi_{m_\psi}(R,Z) < \psi_\times$ as the {\it core}, where $m_\psi$ is the number of grid points required along $\psi$. Here we describe the scheme to generate a flux surface following grid. First, we create a grid on the outer mid-plane starting from grid points at $(R_1+ n \; \Delta R, Z_0)$, where $n \in \{1,2,\cdots, m_\psi\}$, and $\Delta r = (R_{m_\psi}-R_1)/m_\psi,$
such that $\psi(R_1,Z_0) = \psi_1$, $\psi(R_{m_\psi},Z_0) = \psi_{m_\psi}$. Now starting with initial position $(R_i,Z_0)$ we estimate the grid points on the
$\psi = \psi(R_i,Z_0)$ flux surface using the scheme described in Sec.~\ref{Sec:Fluxlines} and Sec.~\ref{Sec:Psiinvariance}. Let $(R_{ij},Z_{ij})$ be the $j$-th grid point along $s$ on the $i$-th flux surface and $\Delta s_{sim}$ be the preferred grid point spacing along $s$, i.e.
\begin{eqnarray}
   \Delta s_{sim} = \sqrt{(R_{ij}-R_{i(j+1)})^2+(Z_{ij}-Z_{i(j+1)})^2}.\nonumber
\end{eqnarray}

\begin{figure}[!ht]
    \begin{center}
   \includegraphics[scale=0.3]{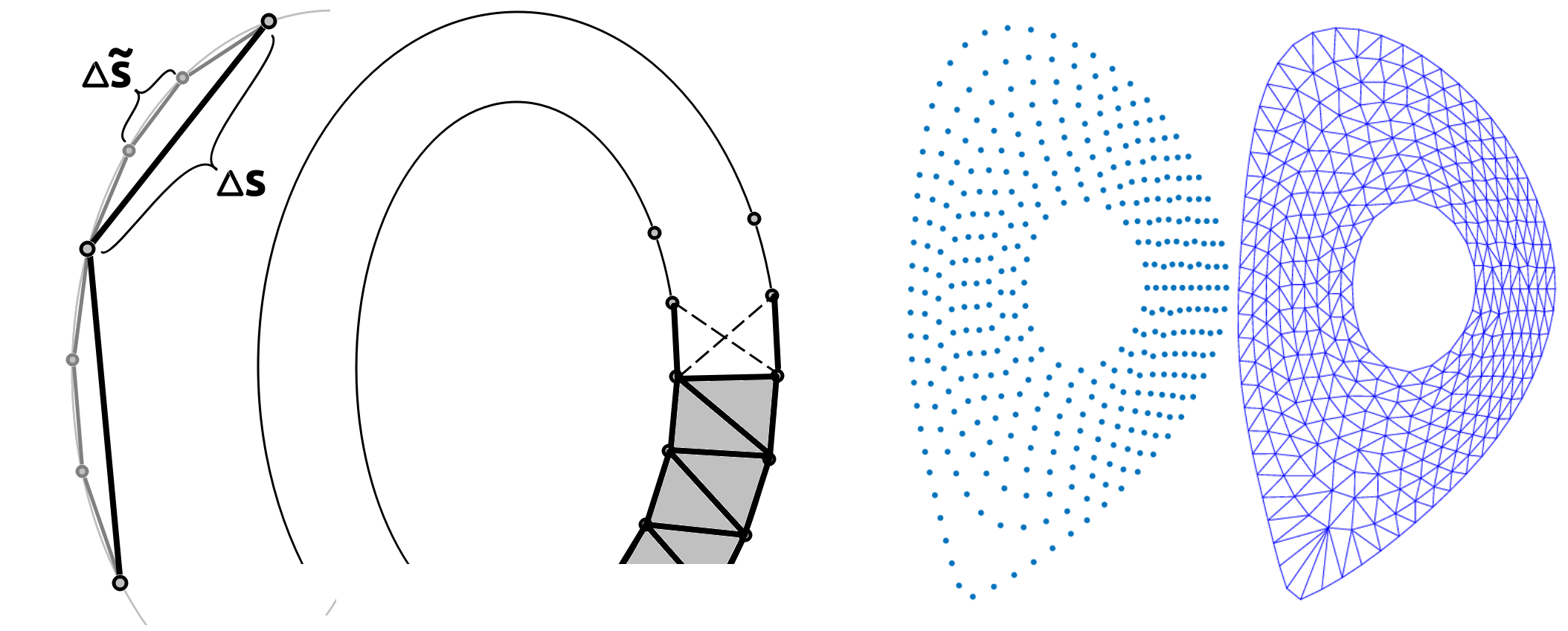}
    {(a)} \hspace{3 cm}  {(b)} \hspace{5 cm}  {(c)} \hspace{3 cm}  {(d)}
    \end{center}
   \caption{(a) Schematic of re-sampling based grid construction process; 
            (b) Construction of triangles between two consecutive
        constant flux surfaces. Note that at every iteration there are two possible
        triangles. The option for which the sum of the lengths of its sides is
        smaller is chosen; (c) Grid generated in the core region $(0.2 
        \psi_\times \leq \psi \leq 0.99 \psi_\times)$, where $\psi_\times$ is the
        flux-function at the X-point, for the DIII-D configuration 
        and the corresponding mesh is shown in (d).}
    \label{fig:Meshing}
\end{figure}

\noindent However, various flux surfaces have different closed-loop lengths and could lead to unevenly spaced first and last grid points on the flux surface. To ensure
that the grids are uniform on the flux surface, we redefine the preferred grid spacing using the following scheme: we first generate a high-resolution grid along $s$ with $\Delta \tilde{s} = (\Delta s_{sim}/\mathbb{F}_s),$
\noindent where $\mathbb{F}_s>1$ (see Fig.\ref{fig:Meshing}(a)). We find the total number
of grid points for a single loop of the $i$-th flux surface, $\tilde{m}_{\theta i}$
and estimate the required number of grid points as, 
\begin{eqnarray}
   m_{\theta i} = \left\lfloor \frac{\tilde{m}_{\theta i}}{\mathbb{F}_s}\right\rceil,
   \label{eqn:mtheta}
\end{eqnarray}
\noindent where $\lfloor \cdot \rceil$ stands for the nearest integer. Then the preferred grid points are $(R_{ij},Z_{ij}) = (R_{i\tilde{j}},Z_{i\tilde{j}}),$ where $\tilde{j} = \left\lfloor (j-1) (\tilde{m}_{\theta i}/m_{\theta i}) \right\rfloor + 1,$ and $\lfloor \cdot \rfloor$ is the floor function. Then the distance between the first grid point and the last grid point of the flux surface is
\begin{eqnarray}
   \Delta \tilde{s} \left(\tilde{m}_{\theta i} - \left\lfloor (m_{\theta i}-1) 
         \frac{\tilde{m}_{\theta i}}{m_{\theta i}} \right\rfloor - 1 \right) 
         = \Delta \tilde{s} \left(\left\lfloor 
          \frac{\tilde{m}_{\theta i}}{m_{\theta i}} \right\rfloor - 1 \right) \nonumber
\end{eqnarray}
and from Eq.~(\ref{eqn:mtheta}), we have
\begin{eqnarray}
   \left( \frac{\tilde{m}_{\theta i}}{\mathbb{F}_s}-1 \right) \leq m_{\theta i} \leq 
         \left( \frac{\tilde{m}_{\theta i}}{\mathbb{F}_s}+1 \right).\nonumber
\end{eqnarray}

\noindent Therefore, the error in the spacing between consecutive grid points is $\sim \Delta \tilde{s}$. Note that in $R_{ij}$ and $Z_{ij}$, the dimension of the $j$ index depends on $i$ and hence is inconvenient to store as matrices. Instead, we define an array $(R_\alpha, Z_\alpha) = (R_{ij},Z_{ij}),$ where, $\alpha = g_i + j$ and $g_i = \sum_{k=1}^{i-1} m_{\psi k}$.

%%%%%%%%%%%%%%%%%%%%%%%%%%%%%%%%%%%%%%%%%%%%%%%%%%%%%%%%%%%%%%%
\subsection{Mesh construction for the core}
\label{Sec:Mesh}

\noindent In the case of closed flux surfaces, as in the core, we triangulate the annular region between the two neighbouring flux surfaces. For each annular region, we start from an edge that connects the two different flux surfaces along the outer midplane. Then moving in the counter-clockwise direction we have two different choices for the triangle as shown in Fig.~\ref{fig:Meshing}. We choose the one with a minimal perimeter and continue this process until all the triangles in the annular region are exhausted. The triangles are stored as a triplet of indices $(\alpha,\beta,\gamma)$ which are the labels/indices corresponding to the 3 vertices that constitute the triangle. Now for each annulus between flux surface indexed $i$ and $(i+1)$, the number of triangles that this approach will construct is $(m_{\theta i} + m_{\theta (i+1)})$. Therefore, the total number of triangles that will be formed at the end of the process will be $n_\Delta=(m_\psi-1)(m_{\theta i} + m_{\theta (i+1)}).$

%%%%%%%%%%%%%%%%%%%%%%%%%%%%%%%%%%%%%%%%%%%%%%%%%%%%%%%%%%%%%%%

\section{Field-aligned interpolation (gather/scatter)}
\label{Sec:Interpolation}

\begin{figure}[!ht]
   \begin{center}
      \includegraphics[scale=0.2]{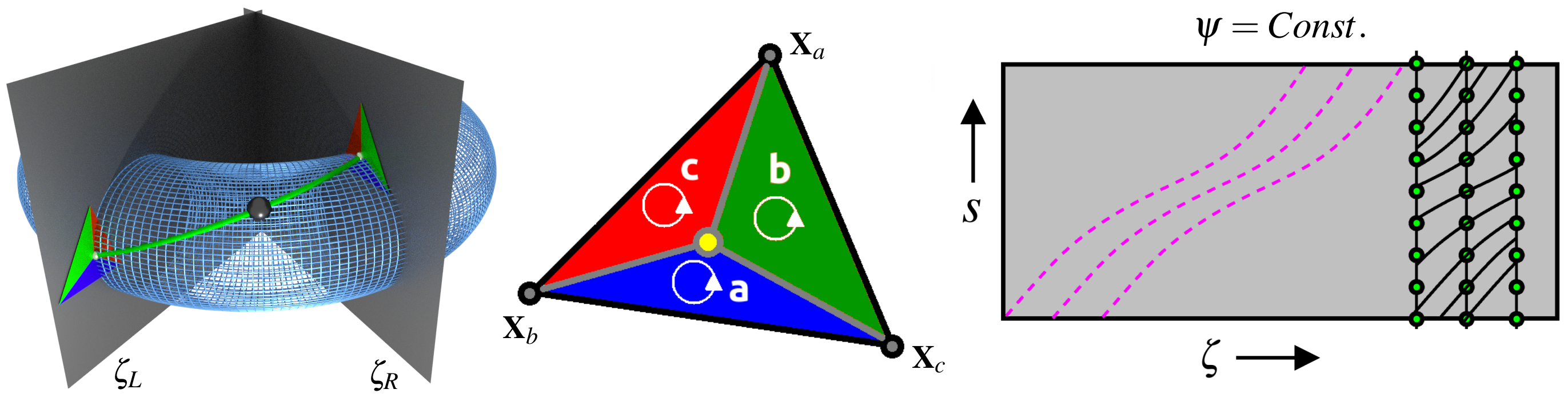} \\
      {(a)} \hspace{4 cm}  {(b)} \hspace{4 cm}  {(c)}
   \end{center}
   \caption{(a) A schematic of $||$-projection along the magnetic field lines on to
   the neighboring left- ($\zeta = \zeta_L$) and right- ($\zeta = \zeta_R$) poloidal mesh.
   The magnetic flux surface is plotted as a mesh only for 
   the visual guide; (b) Area-coordinates visualization of a projected point relative to
   a triangle used for 2d ($\indep$) interpolation in Sec.~\ref{Sec:AreaCoordinates}; 
   (c) A schematic that shows the mesh structure in the $||$-direction on a flux surface,
   where the broken magenta curves represent the magnetic field lines.}
   \label{Fig:Interpolation}
\end{figure}

The presence of a strong applied magnetic field renders the electromagnetic system anisotropic,
and the gyro-centers of charged particles predominantly move along the magnetic field lines.
Furthermore, the typical modes we expect to analyze using the G2C3 have $k_{||} \ll k_{\perp}$.
We use this anisotropy to reduce computational complexity by using a fine grid in the poloidal plane,
$\indep$, and a coarse grid along the $||$-direction.
To perform the first principle, self-consistent simulation we need the grid structure to calculate
the fields using the Poisson solver and transfer data to (gather) and from (scatter) the particles
using the following processes: (i) {\it $||$-projection}: Move along the magnetic field lines
and project the point on the neighbouring poloidal planes (Fig.~\ref{Fig:Interpolation}(a));  
(ii) {\it Triangle locator}: Localize the projected points to within a triangle from the 2d mesh;
(iii) {\it $\indep$-interpolation}: Use area coordinates to perform 2d poloidal plane interpolation
(Sec.~\ref{Sec:AreaCoordinates}, Fig.~\ref{Fig:Interpolation}(b)).

\noindent Within the core region, where every point is uniquely specified by the flux surface and an
angle in the poloidal plane, conventionally (i) is performed via transformation to Boozer
coordinates to reduce the computational cost. But the Boozer coordinates encounter
singularity at the separatrix and fail to extend to the open field line region. We can
circumvent this problem using numerical integration of Eq.(\ref{Eq:FieldLineUpdate}), but it 
is iterative in nature and hence time consuming. Instead, in G2C3 we use a
supervised neural network as a single step proxy integrator, with training data obtained
from numerical integration. Similarly, we adopt the neural network to find the triangle
in (ii). Before we detail the procedures to perform the above operations, first we present
the operation of a neural network as a universal function approximator in 
Sec.~\ref{Sec:FunctionApproximator}. And then in Sec.~\ref{Sec:NNProject} and 
Sec.~\ref{Sec:NNlocate} reformulate (i) and (ii) as a function approximation problem.

\subsection{Neural network as universal function approximator}
\label{Sec:FunctionApproximator}

If $y$ is a multivariate continuous function, then we use a fully-connected vanilla neural
network~\cite{Haykin} to approximate the function in the form,
\begin{eqnarray}
	\tilde{y}_i(\mathbf{x}) := \sum_{j=1}^{N_H} \left[ W2_{ji} \;\; 
              \mathlarger{\mathlarger{\sigma}}\left( \sum_{k=1}^{N_I}
	         W1_{kj} \; x_k + B1_j \right) \right] + B2_i 
	\label{Eq:NN}
\end{eqnarray}
\noindent where $\sigma$ is a non-linear function (we use $\tanh$), referred to as
{\it activation-function}. $N_I$ is the dimension of the multi-variable input $\mathbf{x}$,
$N_H$ is the number of nodes in the hidden layer. $W1$ is a parameter matrix of dimension 
$N_I \times N_H$, and $W2$ of dimension $N_H \times N_O$, where $N_O$ 
is the dimension of the output function 
$\tilde{y}$. And $B1$ is of size $N_H \times 1$ and $B2$ of size $N_O \times 1$.
In general, the neural network acts as a map,
\begin{eqnarray}
    \mathcal{N}: \mathbf{x} \rightarrow \mathbf{y} \nonumber
\end{eqnarray}

\begin{figure}[!ht]
	\centering{
	\includegraphics[scale=0.2]{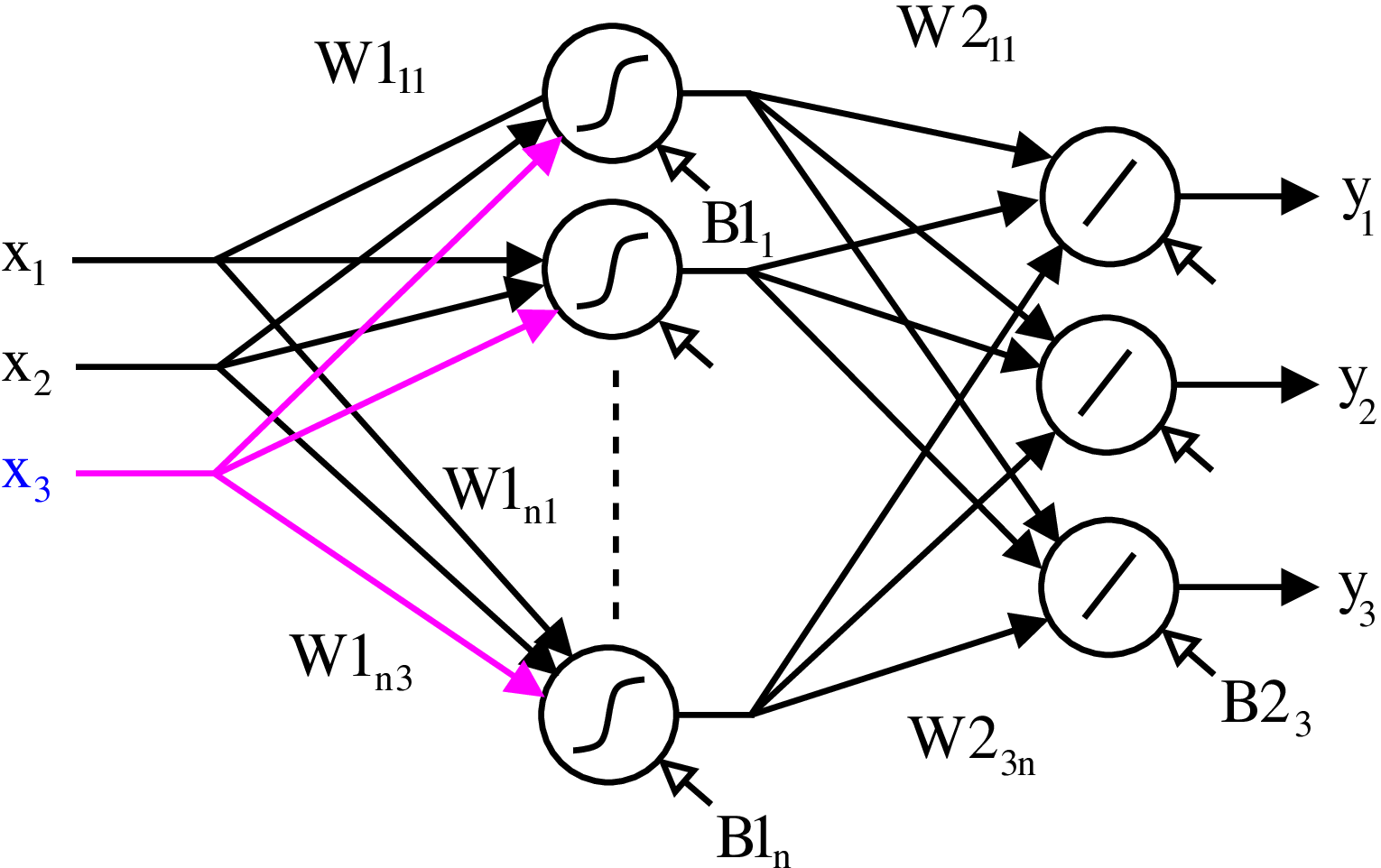}
        \includegraphics[scale=0.3]{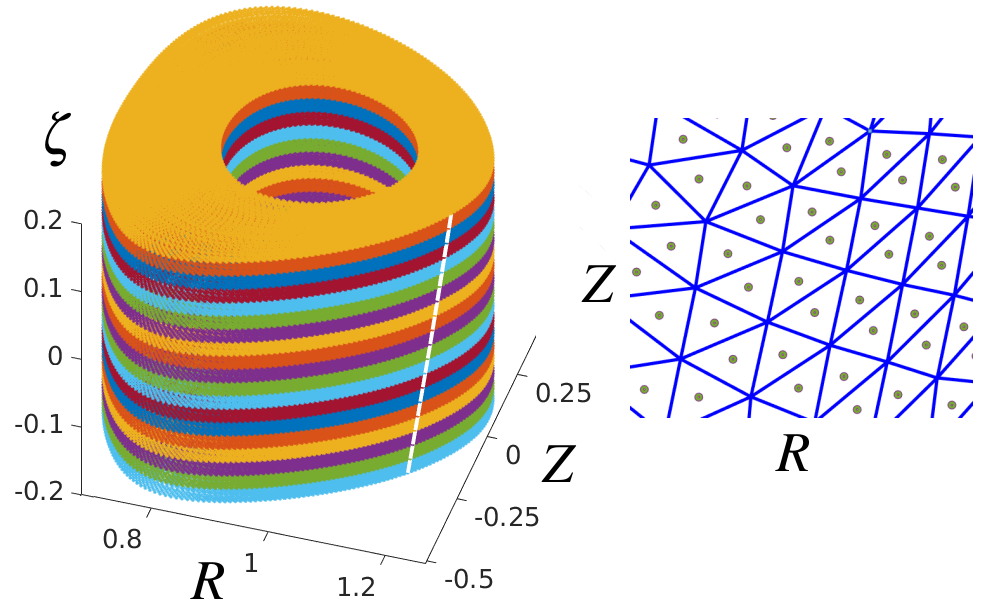} \\
        \hspace{2cm} {(a)} \hspace{6cm} {(b)} \hspace{4cm} {(c)}}
	\caption{(a) Diagrammatic representation of Eq.~\ref{Eq:NN}. Here, $N_I={\color{blue}3}(2)$,
              $N_H=n$, $N_O=3$. The curves within the nodes indicate the form of the
	         activation function, such as the hidden layer has tanh, and the output layer
	         has a linear activation functions. The plot in (b) shows the data points generated
              by solving the ODE in Eq.~\ref{Eq:FieldLineUpdate} for the field
              lines, starting from the grid points. The different colored points indicate discrete
              planes of $\zeta$, and the white curve shows a schematic of a field line. (c) Shows
              the training data for the triangle locator problem, which takes triangle centroid
              coordinates as input, and the corresponding triangle label is the output.}
        \label{Fig:NN}
\end{figure}

\noindent In the mathematical theory of neural networks and approximation theory, the {\it Universal
Approximation Theorem}~\cite{Universal} establishes that such an approximation exists for large enough $N_H$,
and is related to the {\it Kolmogorov–Arnold representation theorem}~\cite{Kolmogorov,Arnold}. Generically, the above
function is diagrammatically represented in Fig.~\ref{Fig:NN}. The network structure, referred to as
the network architecture, is such that it has three layers, namely: (i) input-layer with $N_I$ nodes,
(ii) hidden-layer with $N_H$ nodes, and (iii) output-layer with $N_O$ nodes. The network is fully
connected, as each node of one layer is connected to all the nodes in the next layer. In general,
we can have multiple numbers of hidden layers. The total number of unknowns in the approximating
function is $\#_{parameters} = (N_H N_I) + (N_O N_H) + N_H + N_O$. Here, $N_I$ and $N_O$ are
determined by the input and output dimensions, respectively, and $N_H$ is chosen via convergence
study.

\noindent Given the network architecture, we need to solve for the unknown parameters$\{W1,W2,B1,B2\}$.
Typically, the number of unknowns is large, rendering the problem ill-defined. We will regularize
the problem and formulate the process of determining the parameters as an optimization problem
similar to the curve-fitting problem described in Sec.~\ref{Sec:Training}, referred to as
{\it training}. But before training the network, we process the input and output data for
the best fit.

%%%%%%%%%%%%%%%%%%%%%%%%%%%%%%%%%%%%%%%%%%%%%%%%%%%%%%%%%%%%%%%
\subsubsection{Pre-processing the data:}

\noindent The function $y$ to be approximated is evaluated at finitely many points on the input
domain, such that it densely spans the domain. Each input, $x_i$, is normalized, such that
their numerical values lie in $[0,1]$. Furthermore, the range of output is also standardized, such that
\begin{eqnarray}
	x_k \rightarrow \frac{x_k-x^{min}_k}{x^{max}_k-x^{min}_k} \;\; ; \;\;\;
	y_i \rightarrow \frac{y_i-y^{min}_i}{y^{max}_i-y^{min}_i}\; \nonumber
\end{eqnarray}

%%%%%%%%%%%%%%%%%%%%%%%%%%%%%%%%%%%%%%%%%%%%%%%%%%%%%%%%%%%%%%%
\subsubsection{Training:}
\label{Sec:Training}

\noindent We formulate the problem of estimating the parameters in Eq.~(\ref{Eq:NN}) as an optimization problem, where we need to find the best values of unknown parameters for which the function
\begin{eqnarray}
	\mathcal{L}(\bar{\mathbf{x}},\bar{\mathbf{y}}) = \sum_{i=1}^{N_O} \left[ 
	     \tilde{y}_i(\bar{\mathbf{x}}) - y_i(\bar{\mathbf{x}}) \right]^2,  \nonumber
\end{eqnarray}
referred to as {\it loss-function} is minimum. Now the minimization is performed numerically,
in an iterative form, using the gradient
descent method. Also, notice that, unlike a typical minimization problem, as in
regression analysis, not all the data points are used at a time. The minimization
will be performed one data point at a time, chosen at random, and is called {\it stochastic
gradient-descent}. In general, the parameters $W$ and $B$'s are updated using the gradient descent scheme:
\begin{eqnarray}
	W_{ij} \rightarrow W_{ij} - \eta \; \frac{\partial \mathcal{L}}{\partial W_{ij}}, \;\;\;
	B_i \rightarrow B_i - \eta \; \frac{\partial \mathcal{L}}{\partial B_i}, \nonumber
\end{eqnarray} 
where $\eta \in [0,1)$ is the learning rate. In particular,
\begin{empheq}[right=\empheqrbrace]{align}
   \frac{\partial \mathcal{L}}{\partial B1_i} \; =& \; \left( \sum_{j=1}^{N_O} W2_{ji}
	\frac{\partial \mathcal{L}}{\partial B2_j} \right) \;\;
	\mathlarger{\text{sech}}^2\left( \sum_{k=1}^{N_H} W1_{ik} \; \bar{x}_k + B1_i \right), \nonumber \\
 \frac{\partial \mathcal{L}}{\partial B2_i} \; =& \; 2 \; \left( \tilde{y}_i(\bar{\mathbf{x}})
	   - y_i(\bar{\mathbf{x}}) \right),  \nonumber \\
   \frac{\partial \mathcal{L}}{\partial W1_{ij}} \; =& \; \frac{\partial \mathcal{L}}{\partial 
       B1_i} \; \bar{x}_j,  \nonumber \\
   \frac{\partial \mathcal{L}}{\partial W2_{ij}} \; =& \; \frac{\partial \mathcal{L}}{\partial B2_i} 
	   \;\; \mathlarger{\tanh}\left( \sum_{k=1}^{N_H} W1_{jk} \; \bar{x}_k + B1_j \right).
\end{empheq}

\noindent As per the stochastic gradient descent method, the training data are picked in random sequence, and the above update process is performed. An epoch corresponds to the operation where all the training data has been used once in the updating process. The training process is terminated when the gradients fall below a $0.5 \%$ error cut-off threshold.

\subsubsection{Prediction mode:}

\noindent Once the training is complete, the weights $W$ and $B$'s are fixed. Now Eq.~(\ref{Eq:NN}) gives the approximating function in the analytic form. The same equation can be used to
predict the output for any input data in the trained domain and need not be from the training
dataset. 

%%%%%%%%%%%%%%%%%%%%%%%%%%%%%%%%%%%%%%%%%%%%%%%%%%%%%%%%%%%%%%%
\subsection{\texorpdfstring{$||$}{TEXT}-Projection using neural network}
\label{Sec:NNProject}
%%%%%%%%%%%%%%%%%%%%%%%%%%%%%%%%%%%%%%%%%%%%%%%%%%%%%%%%%%%%%%%

\noindent We use a vanilla neural network, as described in  Sec.~\ref{Sec:FunctionApproximator},
to perform the $||$-projection. 
%Canonically the field lines are obtained by solving the ODEs in Eq.~\ref{Eq:FieldLines}. 
%But the accuracy of the results depends on the scheme used to perform numerical integration, which is invariably iterative in nature and thus computationally expensive. We can overcome these shortcomings 
Given the invariance of the $||$-projection in $\zeta$ in a tokamak, the neural network estimates 
$\delta \vec{\mathbf{X}}$, for a given $\delta \zeta$, as a map
\begin{eqnarray}
   \mathcal{N}_{||}: \left( R, Z, \delta \zeta \right) \longrightarrow \left( \delta R, 
   \delta Z, \delta s \right) \; , \nonumber
\end{eqnarray}
where $R,Z$ belong to the simulation domain and $\delta\zeta \in [-\Delta\zeta,\Delta\zeta]$.

\noindent To generate the training dataset, we estimate the points on the field lines using
the method described in Sec.~\ref{Sec:Fluxlines} thus
constructing the map $\mathcal{N}(R_i,Z_i,n \: \delta\zeta)$, for $n \in \mathbb{Z}$ 
(set of integers),
and $(R_i,Z_i)$ belonging to the simulation grid points on a poloidal plane.

\noindent The input layer has 3 nodes, corresponding to $R$, $Z$, and $\delta\zeta$. We find by
search with multiple of 10 nodes that 30 nodes in the hidden layer is optimal, and the
output layer has 3 nodes corresponding to $\delta R$, $\delta Z$, and $\delta s$. Here
we use the learning rate ($\eta$) of $0.001$ and perform training for $1000$ epochs [cf. Fig.~\ref{Fig:NNEst}(c)].

\noindent Next, we describe how to use a neural network to predict the triangle that encloses
a point in 2d and present the performance of the current neural network in
Sec.~\ref{Sec:NNPerformance}.

%%%%%%%%%%%%%%%%%%%%%%%%%%%%%%%%%%%%%%%%%%%%%%%%%%%%%%%%%%%%%%%%%

\subsection{Triangle locator using Neural network}
\label{Sec:NNlocate}

The mesh generated in the core region (cf. Sec.~\ref{Sec:Mesh}) is partially ordered, and we can label
the triangles using $(i_\psi,i_\theta)$, where $1 \le i_\psi < (m_\psi-1)$ refers to the label of
the constant flux contour, with $m_\psi$ number of flux surface, and
$i_\theta$ refers to the index of the triangle within each annulus. To find the triangle
in which a given point in the poloidal plane lies, we need to find a map 
\begin{eqnarray}
   \mathcal{N}_{\Delta} : (R,Z) \rightarrow (i_\psi,i_\theta)\nonumber
\end{eqnarray}

\noindent But this map is not smooth, as the index $i_\theta$ has a jump across the outer
mid-plane line because of $2\pi$ periodicity. To eliminate the jump we reparametrize the label
$i_\theta$ into two variables $i_{c\theta} = \cos(2\pi i_\theta/n_\psi)$, $i_{s\theta} = 
\sin(2\pi i_\theta/n_\psi)$, where $n_\psi$ is the number of triangles in the annular region.
Hence, our goal is to find a smooth function
\begin{eqnarray}
   \tilde{\mathcal{N}}_\Delta : (R,Z) \rightarrow (\tilde{i}_\psi, \tilde{i}_{c\theta}, \tilde{i}_{s\theta})\nonumber
\end{eqnarray}
such that
\begin{eqnarray}
	i_\psi := \left\lfloor \tilde{i}_\psi \right\rceil \;\;\; \text{and} \;\;\;
	i_\theta :=  \left\lfloor \frac{n_\psi}{2\pi} \arctan( \tilde{i}_{s\theta} , 
		 \tilde{i}_{c\theta} ) \right\rceil  \; , \nonumber
\end{eqnarray}
where $\left\lfloor . \right\rceil$ refers to the roundoff function.

\noindent We use the neural network to estimate the function $\tilde{\mathcal{N}}_\Delta$. 
The input layer has two nodes, corresponding to $R$ and $Z$, and the output layer has three
nodes which correspond to $(\tilde{i}_\psi, \tilde{i}_{c\theta}, \tilde{i}_{s\theta})$. We use
ten nodes in the hidden layer with a learning rate of $\eta = 0.01$, and perform training
for $\sim 1000$ epochs. The training data is obtained by considering the triangle and the coordinates
of the corresponding centroids and tested with randomly distributed points in the mesh.
In the next subsection we present the performance of the neural network as $||$-projector and
triangle locator.

%%%%%%%%%%%%%%%%%%%%%%%%%%%%%%%%%%%%%%%%%%%%%%%%%%%%%%%%%%%%%%%
\subsubsection{Neural network performance analysis:}
\label{Sec:NNPerformance}

\begin{figure}[!ht]
    \centering
    \includegraphics[scale=0.2]{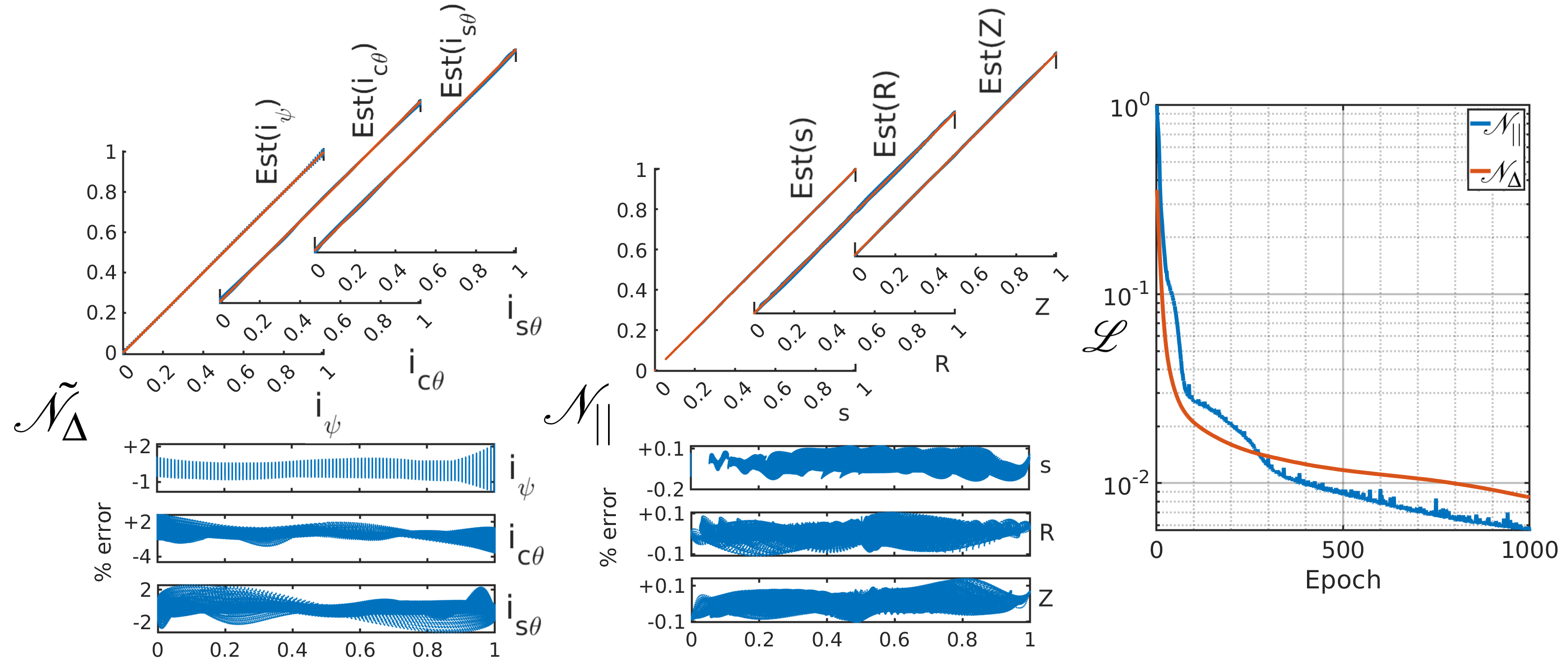}\\
    {\hspace{2cm} (a)} \hspace{5.5cm} {(b)} \hspace{5.5cm} {(c)}
    \caption{ The expected Vs estimated output plots for the three outputs are shown in:
            (a) triangle locator, $\tilde{\mathcal{N}}_\Delta$ (see Sec.~\ref{Sec:NNlocate});
            (b) $||$-projector, $\mathcal{N}_{||}$ (see Sec.~\ref{Sec:NNProject}),
              where the best performance corresponds to the $45^o$ line. The corresponding
              $\%$-errors are shown in the bottom rows. The plot in
              (c) shows the evolution of the loss function during training in the log
              scale for $\mathcal{N}_{||}$ and $\tilde{\mathcal{N}}_\Delta$.}
    \label{Fig:NNEst}
\end{figure}

\noindent We visualize the performance of the trained neural network by generating the
plot of the predicted value versus the expected value in Fig.~\ref{Fig:NNEst}(a) and (b)
corresponding to the three outputs of $\mathcal{N}_{||}$ and 
$\tilde{\mathcal{N}}_{\Delta}$. Figure~\ref{Fig:NNEst}(c) shows the convergence behaviour
of the neural networks using the plot of loss function evolution. In particular,
we find that the network can locate the triangle with a maximum error of three
to four triangle distances from the correct triangle. 

%%%%%%%%%%%%%%%%%%%%%%%%%%

\subsection{Hybrid method for triangle locator}
\label{Sec:Hybrid}

\noindent As indicated in the previous section, the neural network-based triangle locator works well to estimate the triangle in the approximate neighborhood of the correct triangle. The mesh we use in the simulations is partially structured, so we incorporate an iterative local search scheme described in Sec.~\ref{Sec:Iterative} to track down the exact triangle in which the particle lies. G2C3 also has a box-scheme prescribed in~\cite{Zhixin19} as an alternative to the neural network scheme for locating the triangle, which has similar performance to the neural network approach. For the sake of completeness we describe both the approaches below.

%%%%%%%%%%%%%%%%%%%%%%%%%%%%%%

\subsubsection{Triangle locator using box method:}

To perform the 2D poloidal interpolation described above, given a point $(R, Z)$ first we need to find
the triangle it belongs to. To reduce the search space, we first locate the particle within a
rectangular grid in Cartesian $(R,Z)$ space using the box method as,
\begin{eqnarray}
    i_{Box} = \left\lfloor \frac{R-R_{min}}{dR_{Box} \; (R_{max}-R_{min}) }  \right\rfloor, \;\;
    j_{Box} = \left\lfloor \frac{Z-Z_{min}}{dZ_{Box} \; (Z_{max}-Z_{min}) }  \right\rfloor.
    \nonumber
\end{eqnarray}
Here, $(R_{min},R_{max})$ and $(Z_{min},Z_{max})$ refers to the range of $R$ and $Z$ values of the
triangular mesh. The $dR_{Box}$ and $dZ_{Box}$ are parameters that are chosen to be 
$\sim \sqrt{2 \bar{A}_{\Delta}}$, where $\bar{A}_{\Delta}$ is the average area of the triangles.

\noindent Now, we construct a map  
\begin{eqnarray}
    \mathcal{B}:(i_{Box},j_{Box})\rightarrow T  \nonumber
\end{eqnarray}
that takes the given box-index to one of the triangles it encloses. This map helps to find one
of the closest triangles and then we use the iterative scheme described in Sec.~\ref{Sec:Iterative}
to find the correct triangle.

%%%%%%%%%%%%%%%%%%%%%%%%%%%%%%%%%%%%%%%%%%%%%%%%%%%

\subsubsection{Iterative triangle locator using the area coordinates:}
\label{Sec:Iterative}

As shown in Fig.~\ref{Fig:Interpolation}(b), if the point falls within a triangle, then all the
area coordinates are positive. If the point lies outside the triangle, then at least one
area coordinate will be negative. Thus we can build a locator algorithm that uses the area
coordinates to check if the particle is within a triangular element, and use the signs to
find a criterion to move to a neighbouring triangle in the mesh~\cite{Hybrid}.
\begin{figure}[!ht]
   \begin{center}
      \includegraphics[scale=0.2]{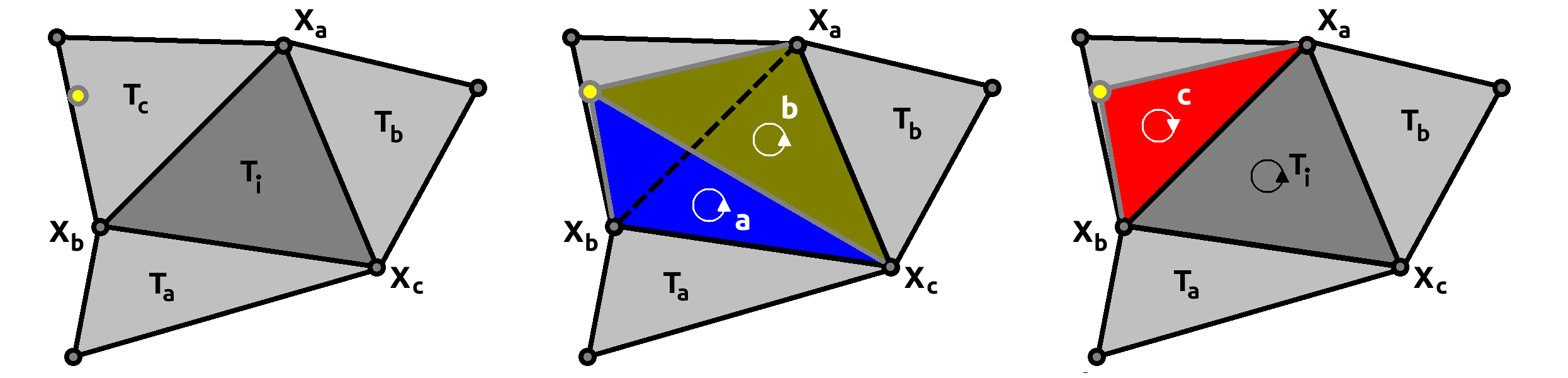} \\
      {(a)} \hspace{4 cm}  {(b)} \hspace{4 cm}  {(c)}
   \end{center}
   \caption{Scheme for locating a particle (yellow) w.r.t. the triangle $T_i$. Here (b)
   and (c) show the area coordinates when the particle is outside the triangle ($T_i$)
   and the arrows indicate the corresponding signs (anti-clockwise is positive). 
   As shown in (b), for this case  $a$-, $b$-coordinates are positive and the 
   $c$-coordinate is negative, as in (c).}
   \label{fig:Tri_Locate}
\end{figure}

\noindent Note that the sign of the area coordinates has the information about on which side of the triangle
the point $\mathbf{X}$ lies. For example, in Fig.~\ref{fig:Tri_Locate} for the case in which the
point lies outside the triangle, $c<0$, implying that $\mathbf{X}$
is farthest from point $\mathbf{X}_c$. So we can shift to the neighbouring triangle element that
shares the corresponding edge. This is accomplished using the triangle label map
\begin{eqnarray}
    \mathcal{T}: T_i \rightarrow (T_a,T_b,T_c)  \nonumber
\end{eqnarray}

\noindent Given a point $(R,Z)$ and an initial guess for the triangle $T_i$, if the corresponding area coordinates are positive, then the point is within the triangle. Else,
if for example $c<a$ and $c<b$, then the triangle label is updated to $T_c$ using $\mathcal{T}$.

\noindent Finally, the hybrid scheme uses the neural network to predict the triangle
corresponding to the particle position and the iterative technique above to correct minor errors. We
find that with enough neural network training, as discussed in Sec  ~\ref{Sec:NNPerformance}, the number of iterative steps required can be reduced
to below four steps for all the particle positions.

%%%%%%%%%%%%%%%%%%%%%%%%%%%%%%%%%%%%%%%%%%%%

\subsection{\texorpdfstring{$\indep$}{TEXT}-interpolation using area coordinates}
\label{Sec:AreaCoordinates}

The 2D triangular grids constructed in Sec.~\ref{Sec:Grids} form the poloidal planes
($\zeta= Const.$) and the mesh sizes are refined to resolve the $k_\perp$.
Now, for any point within a triangle, we can linearly interpolate the quantities
from the triangle vertices to the point using the area coordinates, as shown in 
Fig.~\ref{Fig:Interpolation}(b). Here given a point $(R,Z)$ in the poloidal plane and a triangle 
from the poloidal mesh with coordinates $\mathbf{X}_a$, $\mathbf{X}_b$, and $\mathbf{X}_c$,
which are counter-clockwise in direction, then the corresponding signed areas $a$, $b$, $c$
are:
\begin{eqnarray}
   a = \frac{1}{2 A} \; [ (R_b Z_c - R_c Z_b) + R (Z_b-Z_c) + Z (R_c-R_b) ] \;,\nonumber \\
   b = \frac{1}{2 A} \; [ (R_c Z_a - R_a Z_c) + R (Z_c-Z_a) + Z (R_a-R_c) ] \;,\nonumber \\
   c = \frac{1}{2 A} \; [ (R_a Z_b - R_b Z_a) + R (Z_a-Z_b) + Z (R_b-R_a) ] \;,\nonumber
\end{eqnarray}
where, $A = (1/2)\; [(R_a-R_b)(Z_a-Z_c)-(R_a-R_c)(Z_a-Z_b)]$ is the area of the
triangle, such that $a+b+c=1$. If the point $\mathbf{X}$ lies within the triangle, then
all three areas are positive. But if $\mathbf{X}$ is outside the triangle, then at least
one of the triangle areas will be negative. 

By definition, $\{R, Z\} \; = \; a \; \{R_a, Z_a\} \; + \; b \; \{R_b, Z_b\} \; + \; 
c \; \{R_c, Z_c\}$, and in general for any field $F$ defined on the mesh, $F(R,Z) = a \; 
F_a + b \; F_b + c \; F_c$.

%%%%%%%%%%%%%%%%%%%%%%%%%%%%%%%%%%%%%%%%%%%%%%%%%%%%%%%%%%%%%%%

%%%%%%%%%%%%%%%%%%%%%%%%%%%%%%%%%%%%%%%%%%%%%%%%%%%%%%%%%%%%%%

\subsection{Field gathering (\texorpdfstring{$Grid \; field \rightarrow 
Particle \; field$}{TEXT})}

Finally, we can put together the various processes described in this section to transfer data from
the grid to the particle. For a particle located at $(R,\zeta,Z)$ with $\varphi_L$ and $\varphi_R$
being the data specified on the neighboring poloidal grid, first we $||$-project the points to
$\mathbf{X}_L=(R_L,\zeta_L,Z_L)$ and $\mathbf{X}_R=(R_R,\zeta_R,Z_R)$. The projecting neural
network also estimates the corresponding arc-lengths $s_L$ and $s_R$. Lastly, we locate the
projected points and find the area coordinates $(a_L,b_L,c_L)$ and $(a_R,b_R,c_R)$. Then the
value at the $||$-projected points are
\begin{eqnarray}
    \varphi_* = a_* \; \varphi_{a*} + b_* \; \varphi_{b*} + c_* \; \varphi_{c*},
    \nonumber
\end{eqnarray}
where $*\in \{L,R\}$, such that,
\begin{eqnarray}
    \varphi = \frac{s_R \; \varphi_L + s_L \; \varphi_R}{s_L+s_R}.
\end{eqnarray}
The gathering operation helps us transform any 2D fields, which are defined on the grid points
of poloidal planes, into the 3D field in the bulk of the tokamak compuatational region.

%%%%%%%%%%%%%%%%%%%%%%%%%%%%%%%%%%%%%%%%%%%%%%%

\subsection{Scattering \texorpdfstring{$(Particle\; field \rightarrow Grid\; field)$}{TEXT}:}
\label{Sec:Scatter}

Let the particle position be $(R,\zeta,Z)$. We find the two nearest
poloidal planes, $\zeta = \zeta_L$ and $\zeta = \zeta_R$, such that 
$\zeta_L \le \zeta \le \zeta_R$, and estimate $(\mathbf{X}_L, \mathbf{X}_R)$ 
via $||$-projection. Then the scattering of the field $\varphi$ onto these
points are given by
\begin{eqnarray}
   \varphi_L := \frac{s_R}{s_L+s_R} \; \varphi(\mathbf{X}) ; \;\; \text{and} \; \;
   \varphi_R := \frac{s_L}{s_L+s_R} \; \varphi(\mathbf{X})
\end{eqnarray}
where $s_L$ and $s_R$ are the lengths along the flux-line from $\mathbf{X}_L$ 
and $\mathbf{X}_R$, respectively. Now, we use the triangle locator and area-coordinates 
to scatter the field value onto the nearest grid points in the poloidal plane as:
\begin{eqnarray}
    \varphi_{a*} = a \; \varphi_*, \;\; \varphi_{b*} = b \; \varphi_*, \;\; 
    \varphi_{c*} = c \; \varphi_*, \;\; \nonumber
\end{eqnarray}
where, $* \in \{L,R\}$. In particular, Sec.~\ref{Sec:GKPoisson}
uses scattering operation to estimate the
charge density at the grid from the particle weights.

%%%%%%%%%%%%%%%%%%%%%%%%%%%%%%%%%%%%%%%%%%%%%%%%%%%%%%%%%%%%%%%

%%%%%%%%%%%%%%%%%%%%%%%%%%%%%%%%%%%%%%%%%%%%%%%%%%%%%%%%%%%%%%%
%%%%%%%%%%%%%%%%%%%%%%%%%%%%%%%%%%%%%%%%%%%%%%%%%%%%%%%%%%%%%%%
\section{Particle module (PDE \texorpdfstring{$\rightarrow$}{TEXT} ODEs)}
\label{Sec:Particledynamics}
%%%%%%%%%%%%%%%%%%%%%%%%%%%%%%%%%%%%%%%%%%%%%%%%%%%%%%%%%%%%%%%

In this paper, we focus entirely on the electrostatic-collisionless processes 
with adiabatic electrons to observe the ITG modes. Appropriately, the dynamics
of the ions are described by the Vlasov-Maxwell equations
\begin{eqnarray}
    \frac{\partial f_i}{\partial t} + \dot{\mathbf{x}} \cdot \nabla f_i + 
          \dot{\mathbf{v}} \cdot \frac{\partial f_i}{\partial \mathbf{v}} = 0  \;\;\;\;\; 
    \text{and} \;\;\;\;\;    \nabla \cdot \mathbf{E} = 4\pi \rho\;,
\end{eqnarray}
where $\mathbf{v} = \dot{\mathbf{x}}$, and $f_i(\mathbf{x},\mathbf{v})$ is the ion
distribution function in the 6D phase space. The above set of equations are
solved self-consistently using 
\begin{eqnarray}
    \dot{\mathbf{v}} = \left( \frac{q}{m}\mathbf{E} + \frac{q}{mc} \mathbf{v} \times
            \mathbf{B} \right) \;\;\;\;  \text{and}  \;\;\;\; \rho = q \int \; d^3v \; (f_i-f_e)\;,
\end{eqnarray}
where $\mathbf{B}$ is the applied external magnetic field and $f_e$ corresponds
to the Maxwellian distribution of the adiabatic electron.

\noindent We study micro-turbulence phenomena with time scales that are much larger
than the gyro-motion time scales of ions and electrons, which enables us to integrate
out the gyro-motion. We thus obtain an effective, computationally feasible, 5D
gyro-kinetic phase space, $(R,\zeta,Z,v_{||},\mu)$, and the
corresponding evolution equation is given by~\cite{Catto}
\begin{eqnarray}
   \frac{d}{dt} f_i = \frac{\partial f_i}{\partial t} + \dot{\vec{\mathbf{X}}} \cdot \nabla f_i + \dot{v}_{\parallel} \frac{\partial f_i}{\partial v_{\parallel}} = 0 \;,
\end{eqnarray}
where $f_i$ now is the guiding centre distribution function.

\noindent In a gyro-kinetic PIC simulation, instead of evolving the 5D partial differential
equation (PDE) for $f_i$ we evolve a distribution of $N_i$ marker particles, thus reducing
the problem to $5N_i$ ordinary differential equations (ODE). The next section describes
the ODE's and the corresponding initial conditions.

%%%%%%%%%%%%%%%%%%%%%%%%%%%%

\subsection{Particle push/dynamics}
\label{Sec:ParticlePush}

\noindent The evolution of $f_i$ is captured by the following guiding centre equations of motion
of $N_i$ particles in 5D phase space as: 
\cite{Littlejohn83, Tajinder22}
\begin{eqnarray}
   \dot{\vec{\mathbf{X}}} = v_{\parallel} \frac{\vec{\mathbf{B}}}{B_\parallel^*} + \vec{\mathbf{v}}_E + \vec{\mathbf{v}}_d, \;\;\;\;\;\;\;\;
   \dot{v}_{\parallel} = - \frac{1}{m_i} \frac{\vec{\mathbf{B}}^*}{B_\parallel^*} \cdot \left( \mu \nabla B + Z_i \nabla \phi \right),  
   \label{Eq:ParticleDynamics}
\end{eqnarray}
\noindent and $\vec{\mathbf{B}}^* = \vec{\mathbf{B}} + B v_\parallel/\omega_{ci} (\nabla\times \hat{b})$ is the equilibrium magnetic field at the guiding center position, $B_\parallel^*=\hat b\cdot\vec{\mathbf{B}}^*$, $\vec{\mathbf{v}}_E$ is the $\vec{E}\times\vec{B}$ drift velocity, and $\vec{\mathbf{v}}_d$ is the magnetic drift velocity due to curvature and gradient in magnetic field, which are given by
\begin{eqnarray}
   \vec{\mathbf{v}}_E = \frac{\hat{b}\times \nabla \phi}{B} \;\;\;\; \text{and} \;\;\;\;
   \vec{\mathbf{v}}_d = \vec{\mathbf{v}}_c + \vec{\mathbf{v}}_g = \frac{v_\parallel^2}{\omega_{ci}} \nabla\times \hat{b} + \frac{\mu}{m_i \omega_{ci}} \hat{b}\times \nabla B.
\end{eqnarray}

\noindent Now, if the equilibrium current is suppressed, then
\begin{eqnarray}
 \vec{\mathbf{v}}_d &=& \left( \frac{m_i \; v_\parallel^2 + \mu \; B}{m_i \; \omega_{ci}} \right) \left( \frac{\hat{b}\times \nabla B}{B} \right)
\end{eqnarray}
For an axisymmetric system, components of velocity in cylindrical coordinates are as follows: 
\begin{equation}
   \dot R={v_{||}}\frac{{{B_R}}}{{B_{||}^*}} + \frac{c}{{B_{||}^*}}\left( {\frac{{{B_\zeta }}}{B}\frac{{\partial \phi }}{{\partial Z}} - \frac{{{B_Z}}}{{RB}}\frac{{\partial \phi }}{{\partial \zeta }}} \right) - \frac{{Bv_{||}^2}}{{B_{||}^*{\omega _{ci}}}}\frac{1}{\mathcal{J}}\frac{\partial }{{\partial Z}}\left( {\frac{{R{B_\zeta }}}{B}} \right) + \frac{\mu }{{m_i{\omega _{ci}}}}\frac{{{B_\zeta }}}{{B}}\frac{{\partial B}}{{\partial Z}}, 
\end{equation}
\begin{equation}
  \dot Z={v_{||}}\frac{{{B_Z}}}{{B_{||}^*}} - \frac{c}{{B_{||}^*}}\left( {\frac{{{B_\zeta }}}{B}\frac{{\partial \phi }}{{\partial R}} - \frac{{{B_R}}}{{RB}}\frac{{\partial \phi }}{{\partial \zeta }}} \right) + \frac{{Bv_{||}^2}}{{B_{||}^*{\omega _{ci}}}}\frac{1}{\mathcal{J}}\frac{\partial }{{\partial R}}\left( {\frac{{R{B_\zeta}}}{B}} \right) - \frac{\mu }{{m_i{\omega _{ci}}}}\left( {\frac{{{B_\zeta }}}{{B}}\frac{{\partial B}}{{\partial R}}} \right),
\end{equation}
\begin{align}
    \dot\zeta= {v_{||}}\frac{{{B_\zeta }}}{{B_{||}^*}}\frac{1}{R} + \frac{c}{{B_{||}^*\mathcal{J}}}\left( {\frac{{{B_Z}}}{B}\frac{{\partial \phi }}{{\partial R}} - \frac{{{B_R}}}{B}\frac{{\partial \phi }}{{\partial Z}}} \right) - \frac{\mu }{{m_i{\omega _{ci}}\mathcal{J}}}\left( {\frac{{{B_R}}}{{B_{||}^*}}\frac{{\partial B}}{{\partial Z}} - \frac{{{B_Z}}}{{B_{||}^*}}\frac{{\partial B}}{{\partial R}}} \right)\nonumber\\
    + \frac{{Bv_{||}^2}}{{B_{||}^*{\omega _{ci}}}}\frac{1}{\mathcal{J}}\left[ {\frac{\partial }{{\partial Z}}\left( {\frac{{{B_R}}}{B}} \right) - \frac{\partial }{{\partial R}}\left( {\frac{{{B_Z}}}{B}} \right)} \right],
\end{align}

\begin{gather}
    \dot v_{||}=- \frac{\mu }{m_i}\left[ {\frac{{{B_R}}}{{B_{||}^*}}\frac{{\partial B}}{{\partial R}} + \frac{{{B_Z}}}{{B_{||}^*}}\frac{{\partial B}}{{\partial Z}}} \right] - \frac{{{Z_i}}}{m_i}\left[ {\frac{{{B_R}}}{{B_{||}^*}}\frac{{\partial \phi }}{{\partial R}} + \frac{{{B_\zeta}}}{{RB_{||}^*}}\frac{{\partial \phi }}{{\partial \zeta }} + \frac{{{B_Z}}}{{B_{||}^*}}\frac{{\partial \phi }}{{\partial Z}}} \right]\nonumber\\
    - \frac{{\mu {v_{||}}}}{{m_i{\omega _{ci}}}}\frac{B}{{B_{||}^*}}\frac{1}{\mathcal{J}}\left[ {\frac{\partial }{{\partial R}}\left( {\frac{{R{B_\zeta }}}{B}} \right) - R\frac{\partial }{{\partial Z}}\left( {\frac{{{B_\zeta }}}{B}} \right)} \right]\nonumber\\
    - \frac{{{v_{||}}{Z_i}}}{{m_i{\omega _{ci}}}}\frac{B}{{B_{||}^*}}\frac{1}{\mathcal{J}}\left[ {\frac{{\partial \phi }}{{\partial Z}}\frac{\partial }{{\partial R}}\left( {\frac{{R{B_\zeta }}}{B}} \right) + \frac{{\partial \phi }}{{\partial \zeta }}\left( {\frac{\partial }{{\partial Z}}\left( {\frac{{{B_Z}}}{B}} \right) - \frac{\partial }{{\partial R}}\left( {\frac{{{B_Z}}}{B}} \right)} \right) - R\frac{{\partial \phi }}{{\partial R}}\frac{\partial }{{\partial Z}}\left( {\frac{{{B_\zeta }}}{B}} \right)} \right].
\end{gather}

\noindent If $f_{i0}$ is the equilibrium distribution and $\delta f_i$ is the perturbation ($\delta f_i \ll f_{i0}$), such that $f_i=f_{i0}+\delta f_i$, and the particle weight is defined as $w_i=\delta f_i/f_i$, then
\begin{eqnarray}
 \frac{d}{dt}w_i = \frac{1}{f_{i0}}(1-w_i) \frac{d}{dt} \delta f_i
\end{eqnarray}

\begin{eqnarray}
   \frac{d}{dt} \delta f_i = - \underbrace{\vec{\mathbf{v}}_E \cdot \nabla f_{i0} |_{v_\perp}}_{\text{Drive term}} - \underbrace{\frac{Z_i}{T_i}\left( \hat b\cdot\nabla\phi \right) v_\parallel f_{i0}}_{\text{Parallel term}} - \underbrace{\frac{Z_i}{T_i} \left(\vec{\mathbf{v}}_d \cdot\nabla\phi \right) f_{i0}}_{\text{Drift term}}
\end{eqnarray}

\begin{eqnarray}
   \text{Drive term} &:=& - \vec{\mathbf{v}}_E \cdot \nabla f_{i0} |_{v_\perp} = - \vec{\mathbf{v}}_E \cdot \nabla f_{i0}
                          - \frac{\mu f_{i0}}{T_i} \;\; \vec{\mathbf{v}}_E \cdot \nabla B \nonumber \\
                    &=& - f_{i0} \;\; \vec{\mathbf{v}}_E \cdot \left( \frac{\nabla n_i}{n_i} + A \frac{\nabla T_i}{T_i} \right)\nonumber\\
                    &=& \frac{1}{R B^2} \left[ \left( R B_\zeta \frac{\partial \phi}{\partial Z} - B_Z \frac{\partial \phi}{\partial \zeta} \right) \left(\frac{1}{n_i} \frac{\partial n_i}{\partial R} + \frac{A}{T_i} \frac{\partial T_i}{\partial R}\right) + \right. \nonumber \\
   && \left. \left( B_R \frac{\partial \phi}{\partial \zeta} - R B_\zeta \frac{\partial \phi}{\partial R}\right) \left(\frac{1}{n_i} \frac{\partial n_i}{\partial Z} + \frac{A}{T_i} \frac{\partial T_i}{\partial Z}\right) \right] f_{i0},
\end{eqnarray}
where
\begin{eqnarray}
   A = \left( \frac{2\mu \; B + m_i \; v_\parallel^2 }{2 T_i } - \frac{3}{2} \right).\nonumber
\end{eqnarray}

\begin{eqnarray}
   \text{Parallel term} &:=& - \frac{Z_i}{T_i} \left( \hat b\cdot\nabla \phi \right) v_\parallel f_{i0}
            \nonumber \\
          &=& - \frac{Z_i}{B \; T_i} \left( B_R \frac{\partial \phi}{\partial R} + B_Z 
          \frac{\partial \phi}{\partial Z} + \frac{B_\zeta}{R} \frac{\partial \phi}{\partial \zeta} \right)
          v_\parallel f_{i0}
\end{eqnarray}

\begin{eqnarray}
   \text{Drift term} := - \frac{Z_i}{T_i} \left[ \frac{m_i \; v_\parallel^2 + \mu \; B}{m_i \; \omega_{ci}} \right] \left( \hat b \times \frac{\nabla B}{B} \right) \cdot \nabla \phi f_{i0}\nonumber
\end{eqnarray}
\begin{eqnarray}
   = - \frac{Z_i}{T_i} \left[ \frac{m_i \; v_\parallel^2 + \mu \; B}{m_i \; \omega_{ci}\;B^2} \right] \left[ \left( B_Z \frac{\partial B}{\partial R} - B_R \frac{\partial B}{\partial Z} \right) \frac{1}{R} \frac{\partial \phi}{\partial \zeta} + B_\zeta \frac{\partial B}{\partial Z} \frac{\partial \phi}{\partial R} -  B_\zeta \frac{\partial B}{\partial R} \frac{\partial \phi}{\partial Z} \right]f_{i0}\nonumber\\
\end{eqnarray}

\noindent In the present manuscript we are concerned with the linear simulation
of ITG mode, and the particle trajectory is characterised only by the equilibrium
magnetic field. We use 2nd order Runge Kutta (RK2) method to update the dynamical
quantities in G2C3. Figure\ref{Fig:Domain_Partition}(a) and
(b) show the different banana and passing particle orbits obtained using the
above set of equations for a typical ADITYA-U tokamak discharge.

%%%%%%%%%%%%%%%%%%%%%%%%%%%%%%%%%%%%%%%%%%%%%%%%%%%%%%%%%%%%%%%

\subsection{Particle loading/initialization}

%%%%%%%%%%%%%%%%%%%%%%%%%%%%%%%%%%%%%%%%%%%%%%%%%%%%%%%%%%%%%%%

\subsubsection{\texorpdfstring{$(R,Z,\zeta)$}{TEXT} initialization:}
\label{Sec:ParticleLoading}

\begin{figure}[!ht]
   \begin{center}
   \includegraphics[scale=0.25]{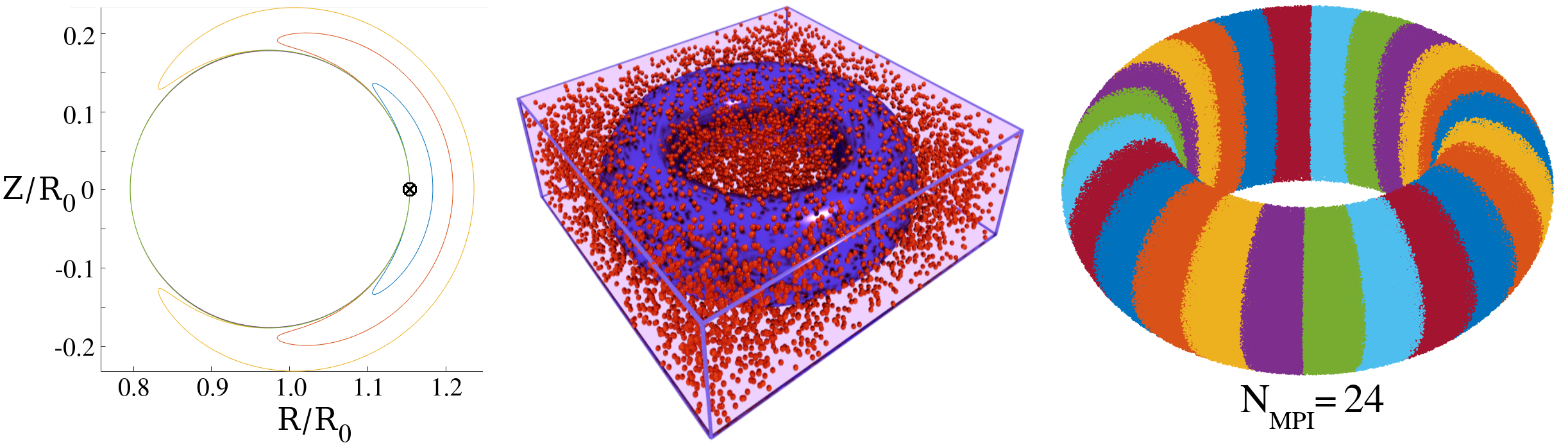}\\
   {(a)} \hspace{4.5 cm} {(b)} \hspace{4 cm} {(c)}
   \end{center}
   \caption{Particle trajectory with different $v_{||}$ starting from the
   same initial position with a transition from banana to passing orbits (ADITYA-U
   Shot $\#$ 36628) are shown
   in (a). The schematic in (b) shows the uniform distribution of particles in
   a box encompassing the toroidal computation domain. The particles belonging
   to the computational domain and the different MPI's are shown with different
   colours in (c). }
   \label{Fig:Domain_Partition}
\end{figure}

\noindent We distribute the particles uniformly in the spatial domain, by generating
a uniform distribution of particles in a 3D Cartesian box and choosing particles which
fall within the computational domain, as shown in Fig.~\ref{Fig:Domain_Partition}(b).
The particles are assigned to different cores of a Message Passing Interface (MPI)
processes so that the computations in
Sec.~\ref{Sec:ParticlePush} can be performed in parallel (See Fig.\ref{Fig:Domain_Partition}(c)).
If we have $N_{MPI}$ number of processors, then each processor is assigned 
$\Delta\zeta = 2\pi/N_{MPI}$ section of the computational domain with $N_i/N_{MPI}$
number of particles each.

\subsubsection{\texorpdfstring{$(v_{||},\mu,w)$}{TEXT} initialization:}

\begin{figure}[!ht]
   \centering{
   \includegraphics[scale=0.45]{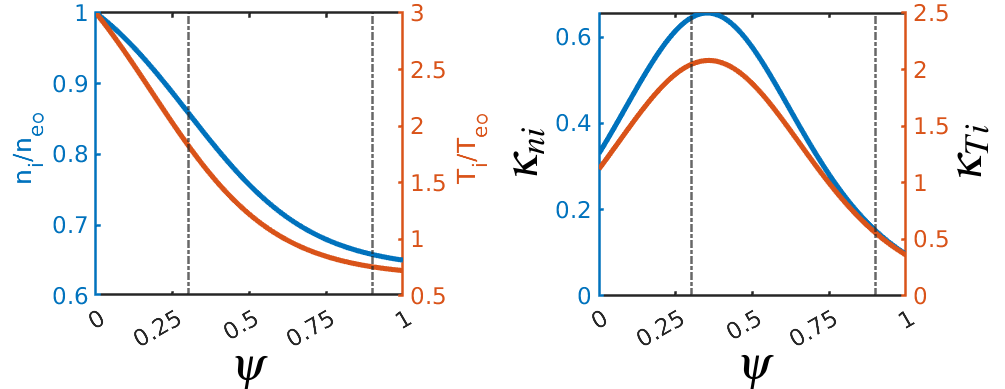}\\
   {(a)} \hspace{6 cm}  {(b)}  \\
   \includegraphics[scale=0.45]{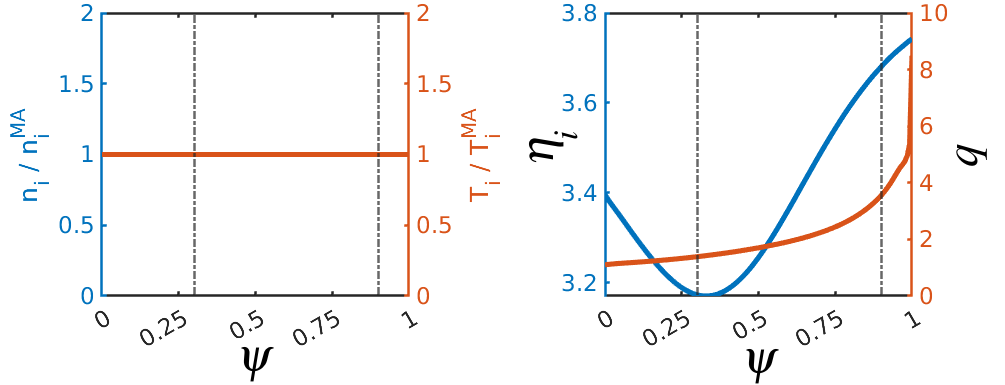}\\
   {(c)} \hspace{6 cm}  {(d)} }
   \caption{In (a) we plot the ion temperature and number density data 
	as a function of normalized flux function. Subplot (c) shows the uniform
	profile of ion temperature and number density we use in the distribution
	function $f(\psi,v_{||},\mu)$; But we retain the profile gradients in
	$\kappa_{Ti} := -\partial( \ln T_i )/\partial\psi$ and 
	$\kappa_{ni} := -\partial( \ln n_i )/\partial\psi$, as shown in (b).
        The plots in (d) show the profile of safety factor
	and the $\eta = \kappa_{Ti}/\kappa_{ni}$. 
	The dotted lines in the plot refer to the range of $\psi$ that belongs
	to the G2C3 computational domain.}
	\label{Fig:PlasmaProfile}
\end{figure}

\noindent The equilibrium phase space guiding centre distribution for ions is
Maxwellian and is given by
\begin{align}
    f_{i0}(|\mathbf{v}|) = f_{i0}(v_{||},\mu) 
    = \frac{n_{i}(\psi)}{\left( \frac{2\pi T_{i}(\psi)}{m_i} \right)^{3/2}} \; 
        exp\left[- \frac{2\mu B + m_i v_\parallel^2}{2 T_{i}(\psi)} \right]
\end{align}
\noindent where $n_{i}$ and $T_{i}$ are the equilibrium radial ion density and temperature profile, respectively. We use the particle loading scheme in which the ion-temperature and ion-density profiles are taken to be constant. In contrast, the gradients in the ion density and temperature profile are retained to drive the instabilities~\cite{XiaoPoP2015}, as shown in Fig.~\ref{Fig:PlasmaProfile}. For this study, we consider the realistic DIII-D geometry shot $\#158103$ at 3050 ms. This discharge is used for resonant magnetic perturbation (RMP) edge localized mode (ELM) \cite{Banerjee21} suppression in the DIII-D pedestal. However, this work uses an experimental equilibrium such that RMPs are not applied and a toroidally symmetric equilibrium is used with the cyclone base case plasma profiles~\cite{Cyclone} to benchmark the ITG mode in the core region. Furthermore, the weights $w$ are initialized to a uniform random distribution in the range of $(-\epsilon,\epsilon)$, where we choose $\epsilon=0.01$.

\subsection{Transfer of particles between MPI's (SHIFTi)}
\label{Sec:shifti}

\begin{figure}[!ht]
   \centering
   \includegraphics[scale=0.3]{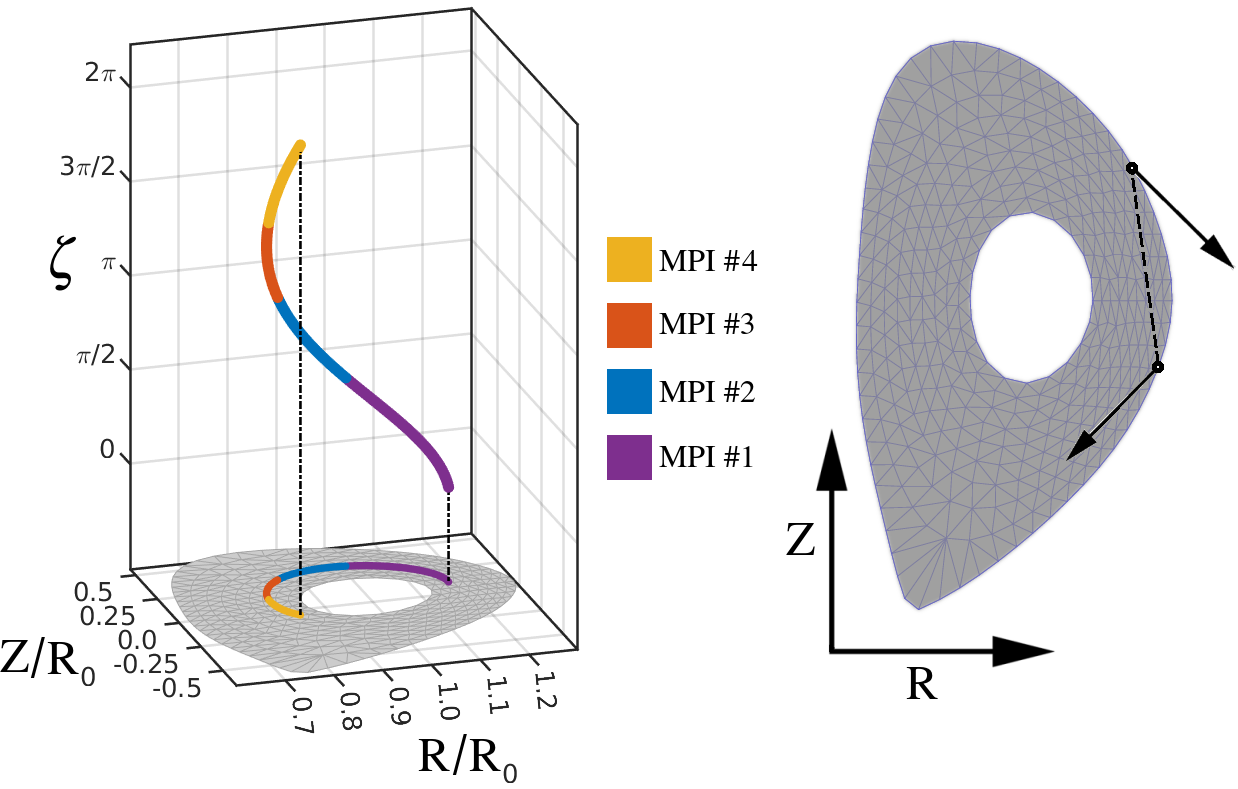}\\
   {(a)} \hspace{4cm} {(b)}
   \caption{(a) Shows the passing particle trajectory shift across different MPIs 
            ($N_{MPI}=4$) during the time evolution; 
            (b) When a particle is lost due to it exiting the simulation region it
	    is reintroduced into the domain, such that 
            energy is conserved. The
            vectors indicate the poloidal component of velocities.}
            \label{Fig:ParticleTrajectory}
\end{figure}

\noindent During the evolution process if a particle crosses from one MPI domain
to the other, the particle's information is transferred to the corresponding MPI
processes.  Figure~\ref{Fig:ParticleTrajectory}(a) shows the trajectory of a set of particles
as they move through different toroidal domains, where the colours indicate the
label of the toroidal domain the particle is assigned to.

\begin{comment}

\begin{figure}[htbp]
\centering
\includegraphics[scale=0.25]{gfx/Shifti.png}
 \caption{A schematic of particle arrays in 3 toroidal domains with corresponding toroidal
           domain labels are indicated by different colors in (a) at current iteration. 
           After the pushing operation the particles may have moved into neighboring MPI's
           as indicated by the color changes in (b). In the next step we rearrange the particles
           as shown in (c) and the MPI communication enable exchange of particle
           data and they are appended to the correct toroidal domains as in (d). And 
           (e) shows the particle shift across different MPI. \textcolor{red}{need to
           track one or two particles to make this figure clear, and provide clear
           description of
           labelling and discarding particles}. \textcolor{blue}{ I will generate
           this figure with lower number of particles, once the CRAY is fully
           operational}.}
 \label{Fig:Shifti}
\end{figure}

\end{comment}

\subsection{Particle conservation and boundary conditions}

\noindent As shown in Fig.~\ref{Fig:ParticleTrajectory}(b), when a particle exits the
computational domain, it is reintroduced again such that the $v_{||in} = v_{||out}$ and
the re-entry point is determined such that the magnetic field strength is conserved,
i.e., $B_{in} = B_{out}$. This ensures that the particle number and the total energy
are conserved.

%%%%%%%%%%%%%%%%%%%%%%%%%%%%%%%%%%%%%%%%%%%%%%%%%%%%%%%%%%%%%%%

%%%%%%%%%%%%%%%%%%%%%%%%%%%%%%%%%%%%%%%%%%%%%%%%%%%%%%%%%%%%%%%
\section{Field Calculations}
\label{Sec:GKPoisson}
%%%%%%%%%%%%%%%%%%%%%%%%%%%%%%%%%%%%%%%%%%%%%%%%%%%%%%%%%%%%%%%

\noindent To complete the PIC cycle for the electrostatic gyrokinetic simulation, we
need to solve the gyrokinetic Poisson's equation. In this section, we describe the
finite element method (FEM) to solve the canonical Poisson's equation in cylindrical
coordinates. Later, we adopt this method to the gyrokinetic case. First, the
particle weight, $w_i$, is scattered to the grid using the method described in 
Sec.~\ref{Sec:Scatter} to obtain the density $\delta n$ at the grid points.
Then we solve for the electric potential as described below.

\subsection{Poisson solver using FEM}
\label{Sec:FEMPoisson}
\noindent In the cylindrical coordinate system, the Poisson's equation is given by
\begin{eqnarray}
    \nabla^2 \phi = \left(\frac{1}{R}\frac{\partial}{\partial R} \left(R \frac{\partial \phi}
        {\partial R}\right) + \frac{\partial^2 \phi}{\partial Z^2} + \frac{1}{R^2} 
        \frac{\partial^2 \phi}{\partial \zeta^2} \right)  = - \delta n
\end{eqnarray}
where $\delta n$ is the density perturbation from quasi-neutral state and $\phi$
is the corresponding electric potential.

\noindent Now, working with a large aspect ratio, such that $\left|\nabla^2
\phi\right| \gg \left|\partial_R \phi / R\right|$ and safety factor, $q>1$,
we have $\partial_{||} \phi \approx \partial_\zeta \phi / R$, and by construction,
$\left|\nabla \phi\right| / \left| \nabla_{||} \phi \right| \sim 
\left(k_\perp/k_{||}\right)$. Therefore,
\begin{eqnarray}
    \nabla^2 \phi \approx
    \nabla^2_\perp \phi \approx \left(\frac{\partial^2 \phi}{\partial R^2} 
           + \frac{\partial^2 \phi}{\partial Z^2} \right) := \nabla_{\hspace{-0.2cm}\indep}^2 \phi
\end{eqnarray}
\noindent which approximates the 3D Poisson's equation to an effective 2D equation on the
poloidal planes. Note that in the effective Poisson's equation, $(R, Z)$ act as Euclidean
coordinates, and there is no coupling between the different poloidal planes. This enables
us to solve for $\phi$ in each poloidal plane on different MPI's with minimal communication.

\noindent We use the FEM to solve for the weak form of Poisson's
equation, given by
\begin{eqnarray}
	\int_\mathcal{A} \chi_\alpha(\mathbf{x}) \; 
        \nabla^2_{\hspace{-0.2cm}\indep} \phi(\mathbf{x})  \; d^2 \mathbf{x} 
	   = - \int_\mathcal{A} \chi_\alpha(\mathbf{x}) \; \delta n(\mathbf{x}) \; d^2 \mathbf{x}, 
    \label{Eq:WeakForm}
\end{eqnarray}
\noindent where $\mathcal{A}$ is the 2D interior domain of the poloidal grid; $\partial\mathcal{A}$
is the boundary of $\mathcal{A}$; $\chi_\alpha(\mathbf{x})$ is a test function,
$\alpha = \{1,2,... N_G\}$, with $N_G$ number of grid points, such that
$\chi_\alpha(\mathbf{x})|_{\partial \mathcal{A}} = 0$, and $\sum_\alpha \chi_\alpha(\mathbf{x}) = 1$
if $\mathbf{x} \in \mathcal{A}$ and $0$ otherwise. Now integrating the LHS by parts we get,
\begin{eqnarray}
	\text{LHS} = \oint_{\partial\mathcal{A}} \chi_\alpha(\mathbf{x}) 
	   \left(\hat{\mathbf{n}}\cdot\nabla_{\hspace{-0.2cm}\indep}\phi(\mathbf{x})\right) ds
          - \int_\mathcal{A} \nabla_{\hspace{-0.2cm}\indep}\chi_\alpha(\mathbf{x}) \cdot \nabla_{\hspace{-0.2cm}\indep} \phi(\mathbf{x})  \; d^2 \mathbf{x},   \nonumber 
\end{eqnarray}
\noindent where $\hat{\mathbf{n}}$ is the outward normal at the boundary. Here we consider only the Dirichlet boundary condition, and by the definition of $\chi$, the first term above is zero.

\noindent Now, we choose the test functions that are localized to the
triangles connected to a vertex and zero outside, i.e. $\chi_\alpha(\mathbf{x}_i)
=\delta_{\alpha i}$, where $\delta$ is the Kroneker delta, for the $i^{th}$ grid point.
Thus the number of test functions is equal to the number of grid points, and each of them
gives rise to a relation in Eq.(\ref{Eq:WeakForm}). The $\alpha^{th}$ integral equation
is restricted to the triangles connected to $\alpha^{th}$ grid point as
\begin{eqnarray}
    \int_\mathcal{A} (.) \; \rightarrow \; \sum^{j,k}_{T_{\alpha jk} \in \{T\}} 
    \int_{T_{\alpha jk}} (.)
\end{eqnarray}
where $T_{\alpha jk}$ is the triangle with vertices $(\alpha,j,k)$; $\{T\}$ is the set
of all triangles in the mesh and the domain of integration is reduced to a triangle. Now,
working in the area coordinates (see Sec.~\ref{Sec:AreaCoordinates}), without loss of
generality we choose $(\alpha,j,k) \rightarrow (a,b,c)$; then $\chi_\alpha(\mathbf{x})
= a(\mathbf{x})$ and the RHS of Eq.(\ref{Eq:WeakForm}) has,
\begin{eqnarray}
    -\int_{T_{\alpha jk}} \chi_\alpha(\mathbf{x}) \; \delta n(\mathbf{x}) \; d^2\mathbf{x}
         = -\int_{T_{abc}} a \; (a \; \delta n_a + b \; \delta n_b + c \; \delta n_c) \sqrt{g} 
         \; da \; db \nonumber \\
         = - \int_{a=0}^1 \int_{b=0}^{1-a} a \; \left( a \; \delta n_a + b \; \delta n_b 
         + (1-a-b) \; \delta n_c \right) \sqrt{g} \; da \; db  \nonumber \\
         = - A_{abc} \left( \frac{2 \; \delta n_a + \delta n_b + \delta n_c}{12} \right)
\end{eqnarray}
and the LHS has,
\begin{eqnarray}
    -\int_{T_{\alpha jk}} \nabla_{\indep} \chi_\alpha(\mathbf{x}) \cdot 
        \nabla_{\indep} \phi(\mathbf{x}) \; d^2\mathbf{x}
         = - \int_{T_{abc}} g^{ef} \; \partial_e a \; \partial_f (a \; \phi_a +
         b \; \phi_b + c \; \phi_c ) \; \sqrt{g} \; da \; db \nonumber \\
         = - \int_{a=0}^1 \int_{b=0}^{1-a} ( g^{aa} (\phi_a - \phi_c) + g^{ab} (\phi_b - \phi_c) )
         \; \sqrt{g} \; da \; db \nonumber \\
         =   - \left(\frac{ R_{bc}^2 + Z_{bc}^2 }{4 \; A_{abc}} \right) \phi_a
           - \left( \frac{R_{ac} R_{cb} + Z_{ac} Z_{cb}}{4 \; A_{abc}} \right) \phi_b
           - \left( \frac{R_{ab} R_{bc} + Z_{ab} Z_{bc}}{4 \; A_{acb}} \right) \phi_c
\end{eqnarray}
where, $R_{ij}:=(R_i-R_j)$, $Z_{ij}:=(Z_i-Z_j)$; $A_{ijk}$ is the (signed) area of
triangle $T_{ijk}$ (see Sec.~\ref{Sec:AreaCoordinates}). The metric tensor is given by,
\begin{eqnarray}
   g_{ij} = \begin{bmatrix}
	   \mathbf{X}_{ab} \cdot \mathbf{X}_{ab} & \mathbf{X}_{ab} \cdot \mathbf{X}_{ac} \\
	   \mathbf{X}_{ab} \cdot \mathbf{X}_{ac} & \mathbf{X}_{ac} \cdot \mathbf{X}_{ac}
   \end{bmatrix} 
   = \begin{bmatrix}
	   R_{ab}^2 + Z_{ab}^2 & R_{ab} \; R_{ac} + Z_{ab} \; Z_{ac} \\
	   R_{ab} \; R_{ac}+Z_{ab} \; Z_{ac}  & R_{ac}^2+Z_{ac}^2
   \end{bmatrix}  \nonumber
\end{eqnarray}
\noindent and $\sqrt{g} = \sqrt{det(g_{ij})} = 2 \; A_{abc}$; Furthermore,
$g^{ij}$ is the matrix inverse of $g_{ij}$.

\noindent Then upon integration over the whole domain $\mathcal{A}$, the contribution at the
$i^{th}$ grid point can be written in a matrix form (with Einstien summation convention)
\begin{eqnarray}
   && \mathbb{K}_{ij} \; \phi_j \; = \; - \mathbb{A}_{ik} \; \delta n_k \; = \; 
            d_i \;\;\;  \nonumber \\
   \Rightarrow \;\;\; && \phi_j \; = \; - \mathbb{K}^{-1}_{ij} \mathbb{A}_{ik} \; 
   \delta n_k \; = \; \mathbb{K}^{-1}_{ij} \; d_i
   \label{Eq:FEM}
\end{eqnarray}
where,
\begin{eqnarray}        
    \mathbb{K}_{ii} = - \sum^{j,k}_{T_{ijk}\in \{T\}} \left(\frac{ R_{jk}^2 + 
            Z_{jk}^2 }{4 \; A_{ijk}} \right), \;\; \;\;
    & \mathbb{A}_{ii} = \sum^{j,k}_{T_{ijk}\in \{T\}} \frac{A_{ijk}}{6},
     \nonumber \\
    \mathbb{K}_{ij} = - \sum^{k}_{T_{ijk}\in \{T\}}
            \left( \frac{R_{ik} R_{kj} + Z_{ik} Z_{kj}}{4 \; A_{ijk}} 
            \right),  \;\;\;\;
    & \mathbb{A}_{ij} = \sum^{k}_{T_{ijk}\in \{T\}} \frac{A_{ijk}}{12},
    \label{Eq:Sparse}
\end{eqnarray}
and for all the boundary grid points, i.e. $i \in \partial \mathcal{A}$ 
\begin{eqnarray}
    \mathbb{K}_{ii} = 1, \;\; \mathbb{K}_{ij} = 0, \;\; \text{if} \;\; j \ne i \nonumber \\
    \text{and} \;\; d_i = \phi_{0i}  \nonumber
\end{eqnarray}
which implies $\phi_i = \phi_{0i}$, where $\phi_{0i}$ is the boundary value, as per
the Dirichlet boundary condition.

\begin{figure}
    \centering{
    \includegraphics[scale=0.2]{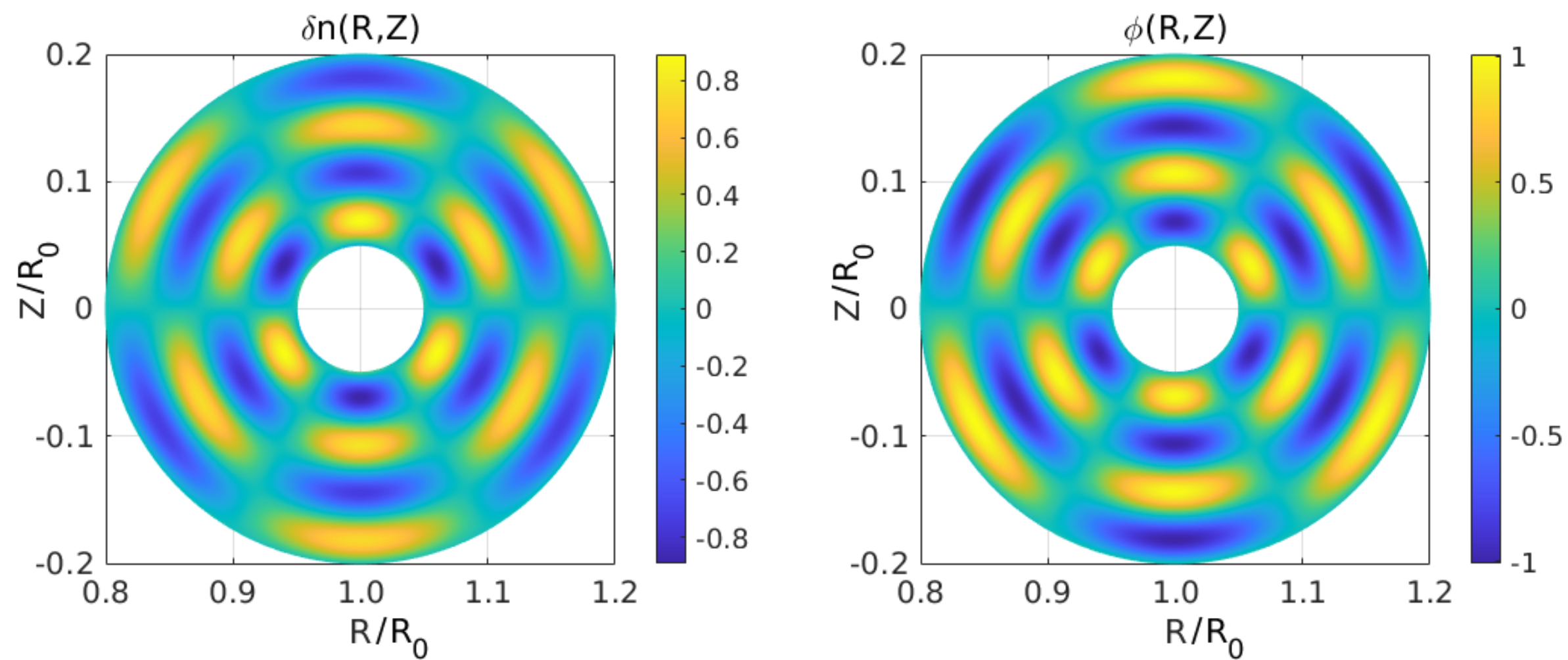} \\
    {(a)} \hspace{6cm} {(b)} \\
    \includegraphics[scale=0.2]{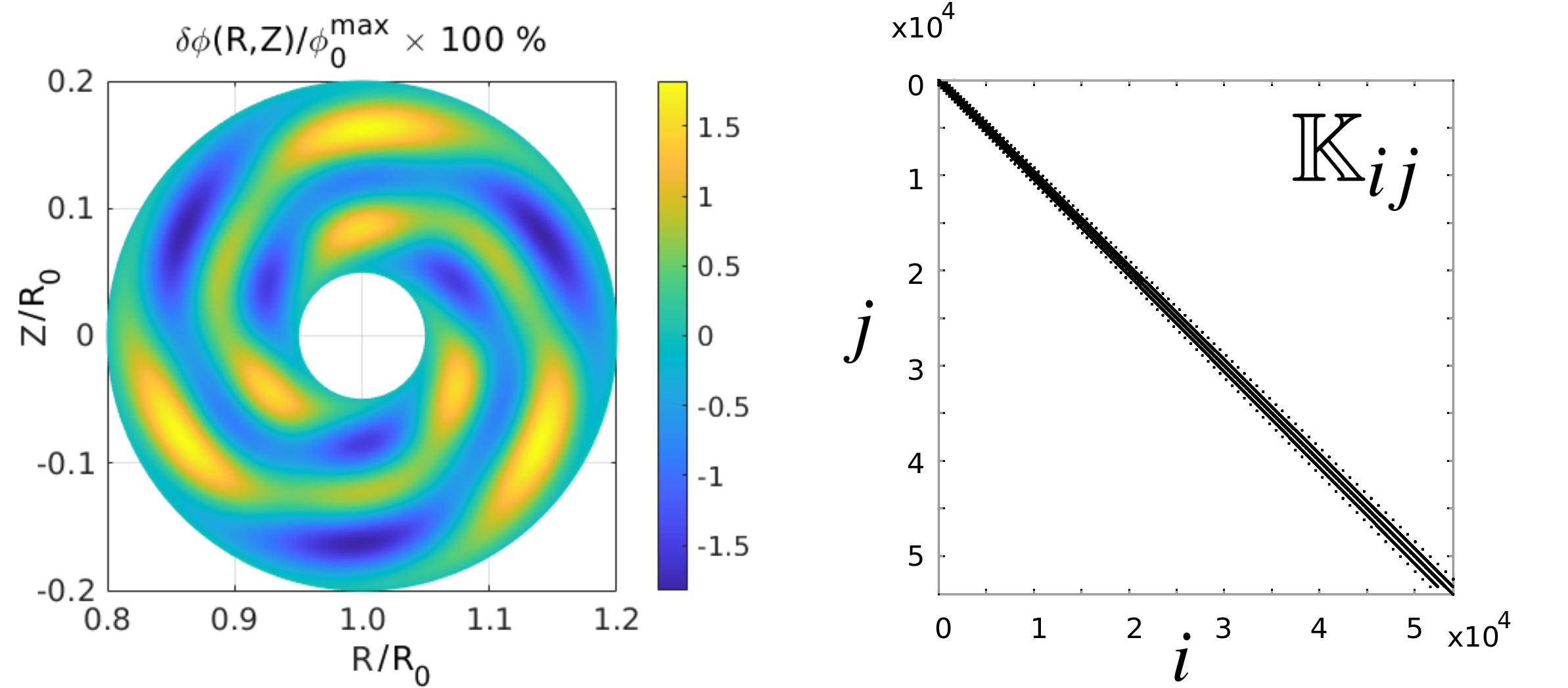} \\
    {(c)} \hspace{6cm} {(d)} }
    \caption{ Plot (a) shows the charge density, $\delta n$, on a circular mesh with a
    hole, with inner radius $r_{min}$ and outer radius $r_{max}$
    and (b) is the estimated electric potential, $\phi$, using the 
    FEM. The difference between the estimated $\phi$ and the analytic solution 
    is given in (c). We get up to $2\%$ accuracy for 50 flux surfaces with
    500 grid points in the angular direction. The binary plot in (d) is the
    visualization of non-zero elements of the sparse matrix $\mathbb{K}_{ij}$.
    Notice that the $\mathbb{K}_{ij}$ matrix gets wider for larger grid indices
    $i$ as the grid points form a spiral starting from the inner circle and have
    an increasing number of grid points in consecutive flux surfaces.}
    \label{Fig:Sparse}
\end{figure}

Notice that a typical $\mathbb{K}_{ij}$ is sparse, as $\mathbb{K}_{ij}=0$ if $T_{ijk} 
\notin \{T\}$ for all $k$ (see
Fig.~\ref{Fig:Sparse}(d)). This enables us to use the MPI-based PETSc package \cite{PETSc}
to compute the resulting sparse-matrix inverse. To test the Poisson solver, we construct
$\mathcal{A}$ with triangular mesh in the form of a circular disc with a hole (see 
Fig.~\ref{Fig:Sparse}). In this case, the polar harmonics are a solution of the form,
$\phi(r,\theta) = \sin(n_\theta \theta) \; \sin\left( \frac{2\pi n_r \bar{r}}{\Delta r} 
\right)$ and $\delta n(r,\theta) = -\nabla^2_{\hspace{-0.2cm}\indep} \phi(r,\theta)$,
where $r=\sqrt{(R/R_0-1)^2+(Z/R_0)^2}$, $\theta=\arctan(Z/(R-R_0))$, $\Delta r = (r_{max}-r_{min})$, 
$\bar{r} = (r-r_{min})$; and $n_r$, $n_\theta$ are the mode numbers in the $r$, $\theta$
directions, respectively. Figure~\ref{Fig:Sparse}(a)-(c) shows the corresponding plots and the estimation
error, for the case $n_r=2$, $n_\theta = 3$.
\begin{comment}
\begin{eqnarray}
\phi(r,\theta) = \sin(n_\theta \theta) \; \sin\left( \frac{2\pi n_r \bar{r}}{\Delta r} 
                    \right) \\
    \delta n(r,\theta) = \frac{\sin(n_\theta \theta)}{r^2 \; \Delta r^2} \left[ 
            2 \pi n_r r \; \Delta r \cos\left( \frac{2\pi n_r \bar{r}}{\Delta r} \right)
            - ( n_\theta^2 \Delta r^2 + 4 \pi^2 n_r^2 r^2 ) \sin\left( \frac{2\pi n_r 
            \bar{r}}{\Delta r} \right) \right]  \nonumber
    \label{Eq:PoissonVerification}
\end{eqnarray}
\end{comment}

%%%%%%%%%%%%%%%%%%%%%%%%%%%%%%%%%%%%%%%%%%%%%%%%%%%%%%%%%%%%%%%
\subsection{Gyrokinetic Poisson's equation}
\label{GKP}

Now, we consider the gyrokinetic case~\cite{Tajjinder23},
\begin{equation}
    \frac{Z_i^2en_{i0}}{T_{i0}}(\phi-\tilde{\phi})=Z_i\delta \bar{n_i}-\delta n_e, 
    \label{eq:GKPoisson}
\end{equation}
\noindent where $\delta \bar{n}_i$ and $\delta n_e$ are the guiding centre
densities of ion and electron, respectively; $n_{i0}$ and $T_{i0}$ are
the equilibrium density and temperature
profiles of ions; $\tilde{\phi}$ is the second gyro-averaged perturbed
potential, defined as
\begin{eqnarray}
    \tilde{\phi}(\vec{\mathbf{x}}) = \frac{1}{2\pi} \int d^3\vec{\mathbf{v}}
    \int d^3\vec{\mathbf{X}} \;\; f_{i0}(\vec{\mathbf{X}}) \; \bar{\phi}(\vec{\mathbf{X}}) \; 
    \delta(\vec{\mathbf{X}} + \vec{\mathbf{\rho}} - \vec{\mathbf{x}})
\end{eqnarray}
where, $\vec{\mathbf{X}}$ and 
$\vec{\mathbf{x}}$ are the position coordinates of the guiding centre and the
particle, 
respectively and $\vec{\mathbf{\rho}}$ is the gyroradius vector. If $\alpha$ is the
gyro angle, then the first gyro-averaged potential $\bar{\phi}(\vec{\mathbf{x}})$ is
defined as
\begin{eqnarray}
    \bar{\phi}(\vec{\mathbf{X}}) = \int d^3 \vec{\mathbf{x}} \; \int 
    \frac{d\alpha}{2\pi} \;\; \phi(\vec{\mathbf{x}}) \; \delta(\vec{\mathbf{x}} - 
    \vec{\mathbf{X}} - \vec{\mathbf{\rho}} )
\end{eqnarray}
 The gyro-averaged physical quantities are calculated
in real space using the four-point method~\cite{Lee87}. Here, each particle is
partitioned into four evenly spaced points on the gyro-ring in a plane perpendicular to
the magnetic field. The gyro-averaging procedure is executed on poloidal planes instead
of gyro-planes since we consider the large aspect ratio cross-section.  The weight of
the particle is uniformly distributed to the 4 gyro-particles and is projected
onto the poloidal planes along the field lines to generate the scattered gyro-charge
and hence the gyro-averaged density. Similarly, the potential estimated using the
gyro-kinetic Poisson solver is gathered onto the 4 gyro-particles and the average
potential is assigned to the particle.

\begin{comment}
\begin{eqnarray}
    \tilde{\phi}(\mathbf{x})=\frac{1}{2\pi}\int d^3\mathbf{v} \int d^3\mathbf{X} \; 
    f_0(\mathbf{X}) \; \bar\phi(\mathbf{X}) \; \delta(\mathbf{X}+\mathbf{\rho}-\mathbf{x}),
\end{eqnarray}
\noindent $\vec{x}$ and $\vec{X}$ are the coordinates of particle position and 
particle guiding center position, respectively. $\bar{\phi}$ is the first-gyro 
averaged perturbed potential given by
\begin{eqnarray}
    \bar\phi(\vec{X})=\int d^3\vec{x}\int \frac{d\alpha}{2\pi} \; \phi(\vec x) \; 
    \delta(\vec{x}-\vec{X}-\vec{\rho}),
\end{eqnarray}
and
\begin{eqnarray}
    \delta \bar{n_i}(\vec{x})=\int d^3\vec{X}\int\frac{d\alpha}{2\pi}\delta f(\vec{X})\delta(\vec{x}-\vec{X}-\vec{\rho}),
\end{eqnarray}
\noindent is the ion perturbed density at the guiding center, and $\alpha$ is the gyro-phase. 
\end{comment}
\noindent For an adiabatic electron model, $\delta n_e=n_{e0}e\phi/T_{e0}$, the above equation can
be rewritten as 
\begin{equation}
    \left(\frac{Z_i^2en_{i0}}{T_{i0}}+\frac{en_{e0}}{T_{e0}}\right)\phi-\frac{Z_i^2en_{i0}}{T_{i0}}\tilde{\phi}=Z_i\delta \bar{n_i}
    \label{eq:GKPoisson2}
\end{equation}
In G2C3, we calculate the second gyro-averaged perturbed electrostatic potential $\tilde{\phi}$
using the Pad\'e approximation~\cite{Hahm}
\begin{eqnarray}
    \tilde{\phi} = \frac{\phi}{1-\rho_i^2 \nabla_{\hspace{-0.2cm}\indep}^2}
\end{eqnarray}

\noindent and the gyro-kinetic Poisson's equation can be written as
\begin{eqnarray}
    && \frac{\phi}{1-\rho_i^2 \nabla_{\hspace{-0.2cm}\indep}^2} = \left( 1 + \frac{1}{Z_i^2}
        \frac{n_{e0} T_{i0}}{n_{i0} T_{e0}} \right) \phi - \frac{1}{e \; Z_i} 
        \frac{T_{i0}}{n_{i0}} \delta \bar{n}_i  \nonumber \\
    \implies &&
    \left\{ F(\mathbf{x}) - \nabla_{\hspace{-0.2cm}\indep}^2 \right\} \phi(\mathbf{x}) 
    = \left\{ G(\mathbf{x}) +
    H(\mathbf{x}) \; \nabla_{\hspace{-0.2cm}\indep}^2 \right\} \delta\bar{n}_i(\mathbf{x})
\end{eqnarray}
where
\begin{align}
    F(\mathbf{x}) = \frac{\left(\frac{1}{\rho_i^2}\right) \left( 
                    \frac{n_{e0}}{T_{e0}}\right)}
                    {Z_i^2 \left(\frac{n_{i0}}{T_{i0}}\right)
                    +\left(\frac{n_{e0}}{T_{e0}}\right)}, \;\;
    G(\mathbf{x}) = \frac{\left(\frac{1}{\rho_i^2}\right)}
                    {Z_i^2 \left(\frac{n_{i0}}{T_{i0}}\right)
                    +\left(\frac{n_{e0}}{T_{e0}}\right)}, \;\;
    H(\mathbf{x}) = \frac{- 1}
                    {Z_i^2 \left(\frac{n_{i0}}{T_{i0}}\right)
                    +\left(\frac{n_{e0}}{T_{e0}}\right)} \nonumber
\end{align}

\noindent The corresponding weak form is 
\begin{align}
    \int_\mathcal{A} \chi_\alpha(\mathbf{x}) \left\{ F(\mathbf{x}) - 
    \nabla_{\hspace{-0.2cm}\indep}^2 \right\} \phi(\mathbf{x}) \; d^2\mathbf{x} = \int_\mathcal{A}
    \chi_\alpha(\mathbf{x}) \; \left\{ G(\mathbf{x})+H(\mathbf{x})\; 
    \nabla_{\hspace{-0.2cm}\indep}^2 \right\} \delta\bar{n}_i(\mathbf{x}) \; d^2\mathbf{x} 
\end{align}

\noindent Now, the first term on LHS can be written as 
\begin{eqnarray}
    F(\mathbf{x}) \; \phi(\mathbf{x}) \approx a(\mathbf{x}) \; F_a \; \phi_a 
        + b(\mathbf{x}) \; F_b \; \phi_b + c(\mathbf{x}) \; F_c \; \phi_c  \nonumber
\end{eqnarray}
such that the discrete version of the weak form at each grid point reduces to
\begin{comment}
    
\begin{eqnarray}
     \frac{1}{4\mathcal{A}_\Delta} \;\; 
    \begin{bmatrix}
        \frac{2}{3} \; F_1 \;  \mathcal{A}_\Delta^2 + x_{23}^2+y_{23}^2 \\
        \frac{1}{3} \; F_2 \;  \mathcal{A}_\Delta^2 + x_{13} \; x_{32} + y_{13} \; y_{32} \\
        \frac{1}{3} \; F_3 \;  \mathcal{A}_\Delta^2 + x_{12} \; x_{23} + y_{12} \; y_{23}
   \end{bmatrix}^T
   \begin{bmatrix}
       \phi_1 \\
       \phi_2 \\
       \phi_3
   \end{bmatrix}
   = \nonumber \\
   \frac{\mathcal{A}_\Delta}{12} \;\; 
    \begin{bmatrix}
        2  \\
        1  \\
        1
    \end{bmatrix}^T 
    \begin{bmatrix}
        G_1 \; \delta\bar{n}_{i1}  \\
        G_2 \; \delta\bar{n}_{i2}  \\
        G_3 \; \delta\bar{n}_{i3}
    \end{bmatrix}  
    - \frac{1}{4\mathcal{A}_\Delta} \;\; 
    \begin{bmatrix}
        x_{23}^2+y_{23}^2 \\
        x_{13} \; x_{32} + y_{13} \; y_{32} \\
        x_{12} \; x_{23} + y_{12} \; y_{23}
   \end{bmatrix}^T
   \begin{bmatrix}
       H_1 \; \delta\bar{n}_{i1}  \\
       H_2 \; \delta\bar{n}_{i2}  \\
       H_3 \; \delta\bar{n}_{i3}
   \end{bmatrix}\nonumber\\
\end{eqnarray}

\end{comment}
\begin{eqnarray}
    \tilde{\mathbb{K}}_{ij} \phi_j = \tilde{d}_i   \nonumber
\end{eqnarray}
where,
\begin{align}
    \tilde{\mathbb{K}}_{ii} = \left( \frac{1}{6} \sum_{T_{ijk}\in\{T\}}^{j,k}  \hspace{-0.2cm}\left( A_{ijk} 
                  \; F_i \right) - \mathbb{K}_{ii} \right), \;\;\;\;\;
    \tilde{\mathbb{K}}_{ij} = \left( \frac{1}{12} \sum_{T_{ijk}\in\{T\}}^{k}  \hspace{-0.2cm}\left( A_{ijk}
                  \; F_j \right) - \mathbb{K}_{ij} \right), \;\; i \ne j   \nonumber
\end{align}
\begin{comment}
\begin{eqnarray}
    \tilde{d}_i =  \left( \sum_{T_{ijk}\in\{T\}}^{j,k} \hspace{-0.2cm}A_{ijk} \left( \frac{2 \; 
        G_i \; \delta\tilde n_i + G_j \; \delta\tilde n_j + G_k \; \delta\tilde n_k}{12} \right) 
           + \mathbb{K}_{ij} \; (H_j\; \delta\tilde n_j) \right)
\end{eqnarray}
\end{comment}
\begin{eqnarray}
    \tilde{d}_i =  \left(  \mathbb{A}_{ij} \; (G_j\; \delta\tilde n_j)
           + \mathbb{K}_{ij} \; (H_j\; \delta\tilde n_j) \right)
\end{eqnarray}
and for all the boundary grid points, i.e. $i \in \partial \mathcal{A}$ 
\begin{eqnarray}
    \mathbb{\tilde{K}}_{ii} = 1, \;\; \mathbb{\tilde{K}}_{ij} = 0, \;\; 
    \text{if} \;\; j \ne i \;\; 
    \text{and} \;\; \tilde{d}_i = \phi_{0i}  \nonumber
\end{eqnarray}
where $\phi_{0i}$ is the boundary value.

%%%%%%%%%%%%%%%%%%%%%%%%%%%%%%%%%%%%%%%%%%%%%%%%%%%%%%%%%%%%%%%

\subsection{Gradient calculation}
\label{Sec:Gradient}

To evaluate the particle dynamics, we need to estimate the gradients
of various fields defined on the grids. We use a FEM based weak formulation of the gradient for 
$\nabla_{\indep}$, where the ${\indep}$ refers to the $(R, Z)$ poloidal plane.
We evaluate $\nabla_{||}$ using the finite difference along the field lines and can be
estimated with higher accuracy than the toroidal component.
This can be estimated as,
\begin{align}
   \nabla_{\parallel} = \hat{b}\cdot\mathbf{\nabla} = \frac{1}{B}\left( 
        \mathbf{B} \cdot \nabla_{\hspace{-0.1cm}\indep} + \frac{B_\zeta}{R} \;
	\partial_\zeta \right)  \;\; \Rightarrow \;\;
   \partial_\zeta = \frac{R}{B_\zeta} ( B \; \nabla_\parallel -
	    \mathbf{B} \cdot \nabla_{\hspace{-0.1cm}\indep} )\nonumber
\end{align}

%%%%%%%%%%%%%%%%%%%%%%%%%%%%%%%%%%%%%%%%%%%%%%%%%%%%%%%%%%%%%%%
\subsubsection{\texorpdfstring{$\nabla_{\hspace{-0.1cm}\indep}$}{TEXT} - component}
\begin{figure}[!ht]
    \centering{
    \includegraphics[scale=0.35]{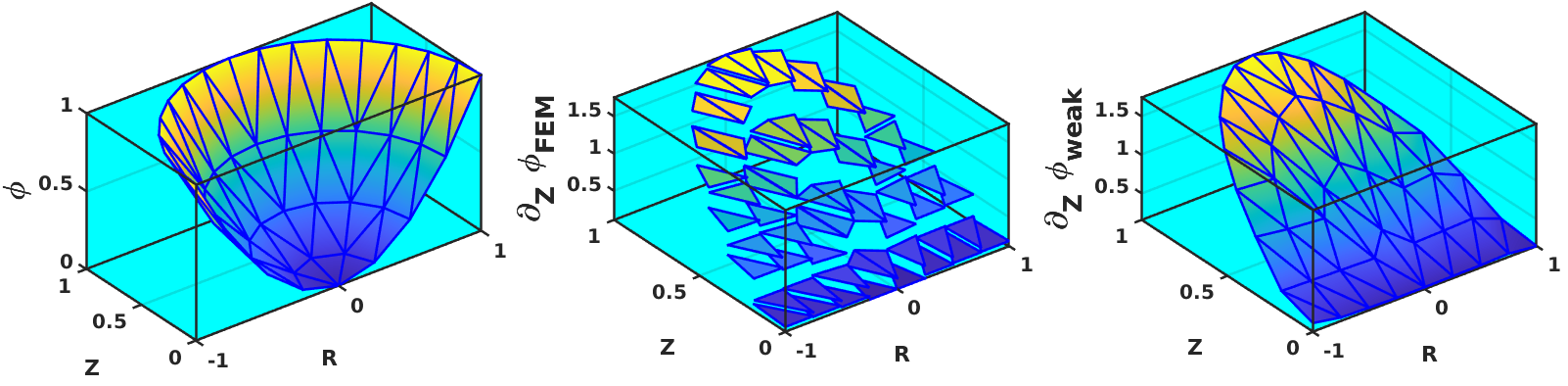} \\
    {(a)} \hspace{5cm} {(b)} \hspace{5cm} {(c)} }
    \caption{Schematic shows (a) the $\phi$-field which we choose to have a parabolic
    form for demonstration. The gradient calculated by direct FEM (which is 
    discontinuous) and the weak form are shown in (b) and (c), respectively.}
    \label{fig:gradient}
\end{figure}

:
The electrostatic potential $\phi$ obtained using the first-order FEM described in
Sec.~\ref{GKP} is a piece-wise linear and continuous function in
$(R,Z)$, given by $\phi(\mathbf{x}) = a(\mathbf{x}) \phi_a + b(\mathbf{x}) \phi_b 
+ c(\mathbf{x}) \phi_c$. So the naive estimate of the electric field,
\begin{eqnarray}
   \mathbf{E}_{\hspace{-0.1cm}\indep} = -\nabla_{\hspace{-0.1cm}\indep} \phi \;\;\;\;
   \Rightarrow \;\; E_{\alpha } = -\partial _{\alpha }\phi, \;\; \alpha\in\{R,Z\} \nonumber 
\end{eqnarray}
is constant within each triangle and is discontinuous across the edges and vertices
(see Fig.~\ref{fig:gradient}). We overcome this problem using the weak formulation
for the gradient, to ensure that the fields do not have a jump when particles shift
across triangles (See Fig.~\ref{fig:gradient}). We have
\begin{eqnarray}
   (a \; E_{\alpha a} + b \; E_{\alpha b} + c \; E_{\alpha c}) \ne - ( \phi_a \; \partial _\alpha a 
        + \phi_b \; \partial_\alpha b  + \phi_c \; \partial_\alpha c ), \nonumber
\end{eqnarray}
but imposing weak equivalence at the $i^{th}$ grid point we get,
\begin{align}
    \int_{T_{abc}} \chi_i \; ( a E_{\alpha a} + b E_{\alpha b} + c E_{\alpha c} ) \sqrt{g} \; da \; db
    = - \int_{T_{abc}} \chi_i \; ( \phi_a \partial _\alpha a + \phi_b\partial_\alpha b  
    + \phi_c \partial_\alpha c ) \sqrt{g} \; da \; db  \nonumber
\end{align}
where $\chi_i$ is the test function, as defined in Sec.~\ref{Sec:FEMPoisson}. Again, choosing
$\chi_i = a$, we get the matrix form as: $\mathbb{A}_{ij} E_{\alpha j} =
- \mathbb{B}_{\alpha ij} \phi _j$, where $\mathbb{A}$ is defined in Eq.(\ref{Eq:Sparse}) and
\begin{comment}
\begin{align}
    \mathbb{A}_{ii} = \sum_{T_{ijk}\in\{T\}}^{j,k} \frac{A_{ijk}}{6},  \;\;\;\;\;\;
    \mathbb{B}_{Rii} = \sum_{T_{ijk}\in\{T\}}^{j,k} \frac{ Z_{jk} }{6}, \;\;\;\;\;\;
    \mathbb{B}_{Rij} = \sum_{T_{ijk}\in\{T\}}^{k} sign(A_{ijk}) \; \frac{Z_{ki}}{6} \nonumber \\
    \mathbb{A}_{ij} = \sum_{T_{ijk}\in\{T\}}^k \frac{A_{ijk}}{12} ,    \;\;\;\;\;\;
    \mathbb{B}_{Zii} = \sum_{T_{ijk}\in\{T\}}^{j,k} \frac{R_{jk}}{6}, \;\;\;\;\;\;
    \mathbb{B}_{Zij} = \sum_{T_{ijk}\in\{T\}}^{k} sign(A_{ijk}) \; \frac{R_{ki}}{6} \nonumber
\end{align}
\end{comment}
\begin{align}
    \mathbb{B}_{Rii} = \sum_{T_{ijk}\in\{T\}}^{j,k} \frac{ Z_{jk} }{6}, \;\;\;\;\;\;
    \mathbb{B}_{Rij} = \sum_{T_{ijk}\in\{T\}}^{k} sign(A_{ijk}) \; \frac{Z_{ki}}{6} \nonumber \\
    \mathbb{B}_{Zii} = \sum_{T_{ijk}\in\{T\}}^{j,k} \frac{R_{jk}}{6}, \;\;\;\;\;\;
    \mathbb{B}_{Zij} = \sum_{T_{ijk}\in\{T\}}^{k} sign(A_{ijk}) \; \frac{R_{ki}}{6} \nonumber
\end{align}
Observe that the $\mathbb{A}$ and $\mathbb{B}$ matrices are sparse and will have similar
structures as in Fig.~\ref{Fig:Sparse}(d).

%%%%%%%%%%%%%%%%%%%%%%%%%%%%%%%%%%%%%%%%%%%%%%%%%%%%%%%%%%%%%%%
\subsubsection{\texorpdfstring{$\nabla_\parallel$}{TEXT} - component}:
We calculate the parallel component of the electric field using the finite difference of potentials
from the two neighbouring poloidal planes, along the field lines. But since this field is a
function of two points on different poloidal planes, the resulting field is located at the
mid-poloidal plane. 
Let $(R_i,\zeta_{mid},Z_i)$ be the $i^{th}$ grid point in the mid-poloidal plane, with the
corresponding $\parallel$-projected points in the neighbouring poloidal plane given by
$\mathbf{X}_L = (R_L,\zeta_L,Z_L)$ and $\mathbf{X}_R = (R_R,\zeta_R,Z_R)$. Then $\zeta_{mid} = 
(\zeta_L+\zeta_R)/2$, and if $s_L$, $s_R$ are the corresponding arc-lengths, then
\begin{eqnarray}
	E_\parallel(R_i,Z_i)|_{\zeta_{mid}} \; = \;
             sign\left( B_\zeta (\zeta_R-\zeta_L) \right) \; 
             \left( \frac{\phi(\mathbf{X}_R) - 
             \phi(\mathbf{X}_L)}{s_R + s_L} \right)
\end{eqnarray}

\begin{comment}

\begin{figure}[!ht]
	\centering{
	\includegraphics[scale=0.4]{gfx/EParallel.png}}
	\caption{Schematic of the different poloidal planes involved in the calculation of
	         $E_\parallel$.}
	\label{Fig:EParallelSchematic}
\end{figure}

\begin{figure}[!ht]
	\includegraphics[scale=0.3]{gfx/EFieldCheck1.png}
	\caption{Testing the $E_\parallel$ calculation. \textcolor{blue}{For the DIII-D case, we will
	have to define $\theta_f$ before using the mode-structure for verification.}}
\end{figure}

\end{comment}

\subsection{Smooth grid quantities}

As a consequence of long-range interactions and only finitely many particles in our
simulations, we encounter spatial noise in the grid quantities, such as density and electric
potential. We improve the SNR (signal-to-noise ratio) by performing the smoothing operation on the
poloidal plane using a kernel of the form $\Phi_i \rightarrow \mathbb{S}_{ij} \Phi_j$, where
\begin{eqnarray}
    \mathbb{S}_{ii} = \frac{1}{2} \; ; \;\;\;\; \mathbb{S}_{ij} = \frac{1}{2 N_i}, 
    \;\; \text{with} \;\; 
    N_i = \hspace{-0.3cm}\sum_{T_{ijk}\in \{T\}}^{j,k} \hspace{-0.3cm} 1 
\end{eqnarray}
where $N_i$ is the number of nearest neighbour grid points to $i$. 

The number of grid points in the $\parallel$-direction is small compared to the ${\indep}$-
directions, as $k_{\parallel} \gg k_\perp$, which adds to the noise. So, we perform a
$\parallel$ smoothing as well by averaging the data from the neighbouring poloidal planes
along the field lines. This concludes our discussion of the primary modules for an
electro-static PIC simulation.

\section{Simulation of the linear ITG mode}
\label{Sec:ITG}

We have so far described all the modules needed to perform a gyrokinetic
electrostatic PIC simulation. In this section, we proceed to benchmark
our implementation by studying the Ion Temperature Gradient (ITG) driven
instability modes. For this study, we use the equilibrium magnetic
field obtained from the DIII-D shot $\#158103$, from a g-file generated
using EFIT~\cite{LaoEFIT90}, with cyclone profile. 

We follow the below-mentioned setup sequence and parameters
to initialize the simulation:
\begin{itemize}
    \item Grid generation (Sec.~\ref{Sec:Equilibrium}): We perform the simulation in an
    annular region ($\psi_1/\psi_\times=0.3$ and $\psi_{m_\psi}/\psi_\times=0.9$).
    We use $100$ flux surfaces ($m_\psi$) with $32$ poloidal planes ($m_\zeta$),
    such that the radial and angular resolution in the poloidal planes are 
    $\Delta r/R_0 \sim 2.5 \times 10^{-3}$ and $\Delta s/R_0 \sim 5 \times 10^{-3}$,
    respectively. Note that we have chosen the grid sizes such that 
    $\Delta r/\rho_i\approx 2 $, to resolve the poloidal components.
        
    \item Train the neural network for $\parallel$-projection and triangle-locator
        (Sec.~\ref{Sec:Interpolation}): We use the $2$ layered neural network to find the maps
        $\mathcal{N}_\parallel$ and $\mathcal{N}_\Delta$ which are trained with
        a dataset of size $7.4 \times 10^5$ and $7.4 \times 10^4$, respectively.
        We use a stochastic gradient descent-based optimizer, with a quadratic
        loss function. Upon training, for the $\mathcal{N}_\parallel$ neural
        network we observe that the loss converges to $\sim 1 \%$ and for the
        $\mathcal{N}_\Delta$ network the predicted triangle falls within the
        next nearest neighbour (which is corrected using the iterative scheme
        in Sec.~\ref{Sec:Hybrid}). The loss evolutions of the networks are given in
        Fig.~\ref{Fig:NNEst}.
        
    \item Load profiles (Sec.~\ref{Sec:ParticleLoading}): We load the profile
        data shown in Fig.~\ref{Fig:PlasmaProfile} onto the simulation grid. The
        ion and electron 
        temperatures at the magnetic axis are $5.068$ keV and $1.690$ keV,
        respectively. We consider hydrogen ions $(Z_i=1)$, and hence by the 
        quasi-neutrality requirement we have $n_i=n_e$, equal to 
        $3.28 \times 10^{13} cm^{-3}$ at the magnetic axis.
        The purpose of our study is to benchmark the capabilities
        of G2C3 to handle realistic geometry with minimum technical
        complexity. The fully self-consistent treatment of the
        density/temperature proﬁle and the equilibrium will be addressed in
        subsequent work.
        
    \item Load particles (Sec.~\ref{Sec:ParticleLoading}): We load 
        $32\times 10^6$ number of particles, uniformly into the simulation
        domain, with Maxwellian velocity distribution. We initialize the
        particle weights with uniform distribution, such that 
        $w \in [-0.005,0.005]$.
        
    \item Initialize the FEM matrices (Sec.~\ref{Sec:GKPoisson}): We initialize the
        $\mathbb{K}_{ij}$, $\mathbb{A}_{ij}$, $\mathbb{B}_{Rij}$, and
        $\mathbb{B}_{Zij}$ as sparse matrices to estimate the electric
        field. We apply the Dirichlet boundary condition and enforce the
        field quantities to vanish at the boundaries. The sparse
        matrices are partitioned across multiple MPIs
        and handled by the PETSc package~\cite{PETSc}. We use a combination of a
        Krylov subspace method~\cite{Krylov} and a preconditioner~\cite{PETSc} to
        perform iterative numerical inversion to find the solution of
        the linear system, with a tolerance level of $\sim10^{-7}$.
    
\end{itemize}

We perform a linear adiabatic-electron electrostatic gyrokinetic simulation,
with a time step size of $0.02 \; R_0/C_s$,
where $R_0$ is the major radius and $C_s$ is the ion sound speed.
Now, the system is run through the main PIC loop, namely:
scatter particle weights to the grid (Sec.~\ref{Sec:Interpolation});
estimate the potential on the grid (Sec.~\ref{Sec:GKPoisson});
estimate the electric field (Sec.~\ref{Sec:Gradient});
gather the electric field to the particle (Sec.~\ref{Sec:Interpolation});
push the particle (Sec.~\ref{Sec:Particledynamics});
transfer particles to their respective MPI (Sec.~\ref{Sec:shifti}).
The above sequence of operations is repeated
iteratively to evolve the mode structure growth.

\begin{figure}[!ht]
    \centering
    \includegraphics[scale=0.37]{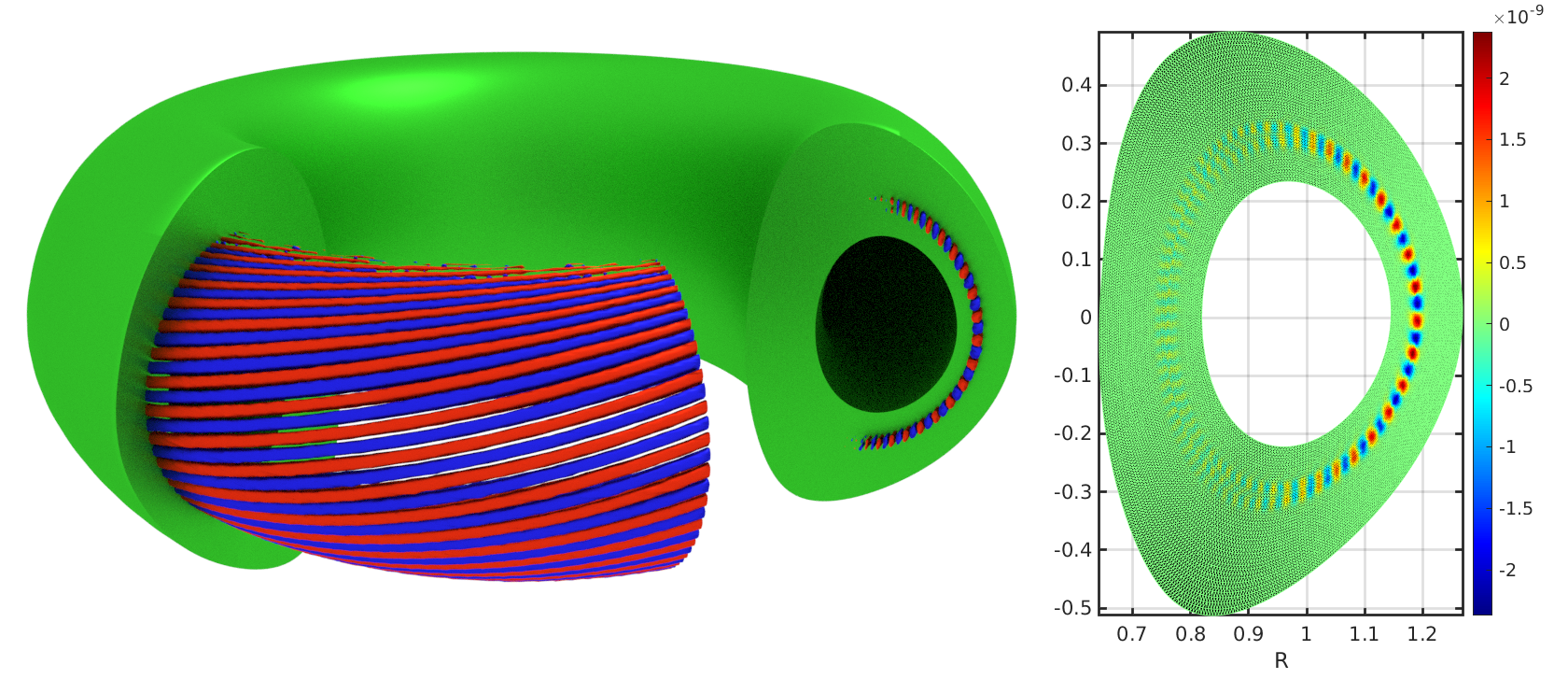}
    \caption{In (a) we show the iso-surface plot of the electric
    potential $\phi(R,\zeta,Z)$ for the ITG mode, which demonstrates
    the twisted nature of the structure; and (b) shows the
    corresponding poloidal cross-section, $\phi(R,Z)$ at $\zeta=0$.}
    \label{Fig:ModeStructure}
\end{figure}

We simulate for $60 \; R_0/C_s$ time to find an ITG mode signature in
the electrostatic potential on the poloidal plane shown in
Fig.~\ref{Fig:ModeStructure}. The ITG mode becomes unstable due to the ion
temperature gradient in the core. The linear eigenmode
has a typical ballooning mode structure and is localized on
the outer mid-plane side where the curvature is bad.
The next section details
the transformation to Boozer coordinate to extract the mode
numbers.

\subsection{Mode structure analysis}

To analyze the mode structure harmonics in the core we transform to
the Boozer coordinates~\cite{Boozer}, such that each point on the poloidal plane
can be represented in terms of the flux function and an angle variable as,
\begin{eqnarray}
    \mathcal{T}_{Boozer}: (R,\zeta,Z) \rightarrow (\psi,\theta,\zeta)
\end{eqnarray}
where the field lines on each flux surface are represented by 
$\bar{\theta} = q(\psi) \; \zeta + \theta$, with $q(\psi)$ as the safety
factor. To find $\theta$, we move along the field line in the $\hat{b}$
direction until we cross the outer mid-plane at $(R^+,\zeta^+,0)$, and
move in the $-\hat{b}$ direction to find $(R^-,\zeta^-,0)$. Then 
\begin{eqnarray}
   q = \frac{2\pi}{\left( \zeta^+-\zeta^- \right)} 
   \;\;\;\; \text{and}  \;\;\;\;
   \theta = \frac{2\pi \; \left| \zeta^- \right|}{\left( \zeta^+ 
   - \zeta^- \right)}.
\end{eqnarray}

An $(m,n)$-mode is of the form $\sim A_{m,n} \; sin(m \; \theta
- n \; \zeta)$. We use 2D Fast-Fourier-Transform (FFT)~\cite{FFT} to 
perform the $(\theta,\zeta) \mapsto (k_\theta,k_\zeta)$ transformation
on each flux surface. To calculate the FFT we construct a square grid
in the $(\theta,\zeta)$ coordinate for each flux surface, with
$N_{max} \times N_{max}$ grids, where $N_{max}$ is the number of
grid points in a poloidal plane on the outermost flux surface. The
$\mathcal{N}_\parallel$-map is used to perform field-line-aligned interpolation
to transfer the field from the computational grid to the Boozer grid.

\begin{figure}[!ht]
    \centering{
    \includegraphics[scale=0.32]{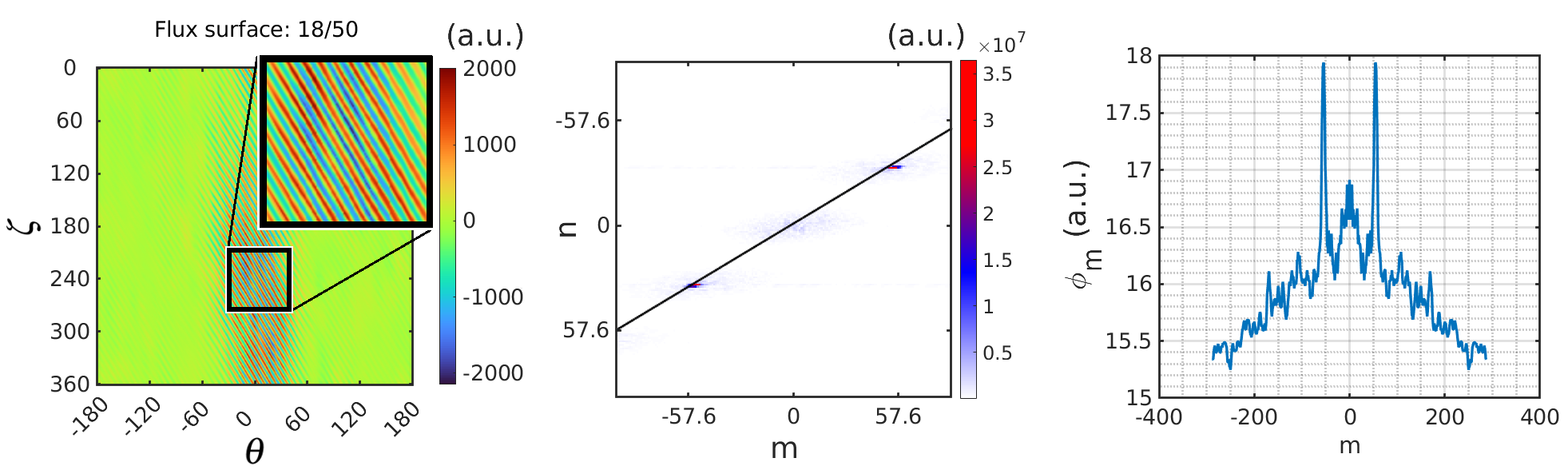} \\
    \hspace{1cm} {(a)} \hspace{5cm} {(b)} \hspace{5cm} {(c)} }
    \caption{(a) The $\phi$-field as a function of $(\theta,\zeta)$ on
    the flux-surface with the highest amplitude ($\psi_{max}$). The
    inset shows the linear structure of the ITG mode; (b) FFT of the
    $\phi$-field in (a). The peaks in amplitude correspond to the
    $(m,n)$ (and $(-m,-n)$) value
    of the mode, and we note that $m/n (1.7188) \approx q(\psi_{max})$
    as indicated by the black line;
    (c) We sum over all the $n$-modes to get the 1D plot, from which we
    can identify the $m$-value of the ITG mode.}
    \label{Fig:ModeAnalysis}
\end{figure}

\begin{figure}[!ht]
    \centering{
    \includegraphics[scale=0.32]{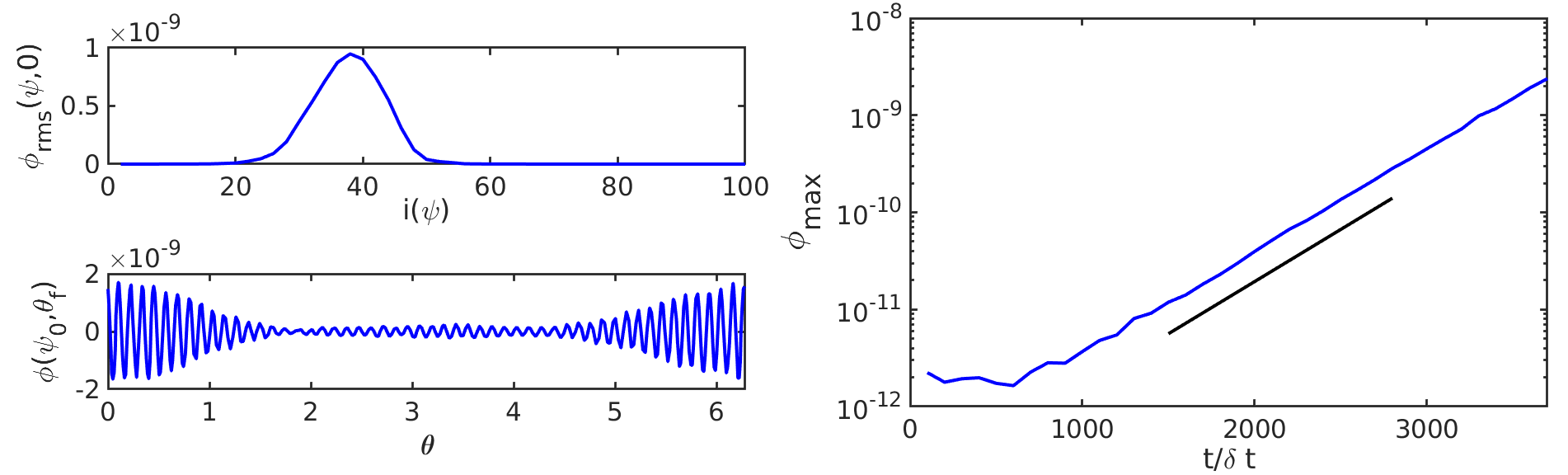} \\
    \hspace{2cm} {(a)}  \hspace{7cm} {(b)} }
    \caption{We calculate the rms value of the $\phi$-field over each
    flux-surface (outer mid-plane, $Z=0$), to obtain the profile shown in
    the top panel of (a), which shows the localization of the ITG mode.
    In (a) bottom panel, we plot the variation in $\phi$
    w.r.t. the Boozer angle on the flux surface with maximum amplitude.
    The oscillations indicate the $m$-value and
    the modulation is such that the instability amplitude is highest in
    the outer mid-plane (bad curvature region). We plot the maximum value of
    $\phi$ in (b) as a function of time to find an exponential growth
    in the mode strength.}
    \label{Fig:Growth}
\end{figure}

Figure~\ref{Fig:ModeAnalysis}(a),(b) shows the $\phi$-field and the
corresponding Fourier representation for the obtained ITG mode on the
flux surface where the amplitude is maximum. Now, to extract only the $(m,n)$
mode we build the filter response function in the Fourier space as
\begin{eqnarray}
	\mathcal{F}(k_\theta,k_\zeta) := \delta_{k_\theta,m} \; \delta_{k_\zeta,n}
\end{eqnarray}
where the $\delta$'s are the Kroneker delta functions. Then the filtered
output is given by
\begin{eqnarray}
	\bar{\Phi}(k_\theta,k_\zeta)|_\psi = \Phi(k_\theta,k_\zeta)|_\psi \; 
	             \mathcal{F}(k_\theta,k_\zeta)
\end{eqnarray}
which is in the Fourier space, and we perform an inverse FFT to finally
obtain the filtered potential in $(\theta,\zeta)$ space. In linear simulations,
we need to ensure that the noise does not excite other modes in the system
which is achieved using the filter operation. Using the mode structure analysis
described above we extract profiles of the dominant $(m,n)$ mode as shown
in Fig.~\ref{Fig:Growth}(a),(b) and the corresponding exponential amplitude growth
in  Fig.~\ref{Fig:Growth}(c). Furthermore, we extract the growth rates of the
sub-dominant $n$-modes to generate the dispersion plot in 
Fig.~\ref{Fig:Dispersion}(a), which shows a peaks at $n=32$ and gradually
falls off away from it.

%-----------------------------------------------------------------------

\subsection{Convergence analysis:}

\begin{figure}[!ht]
    \centering{
    \includegraphics[scale=0.37]{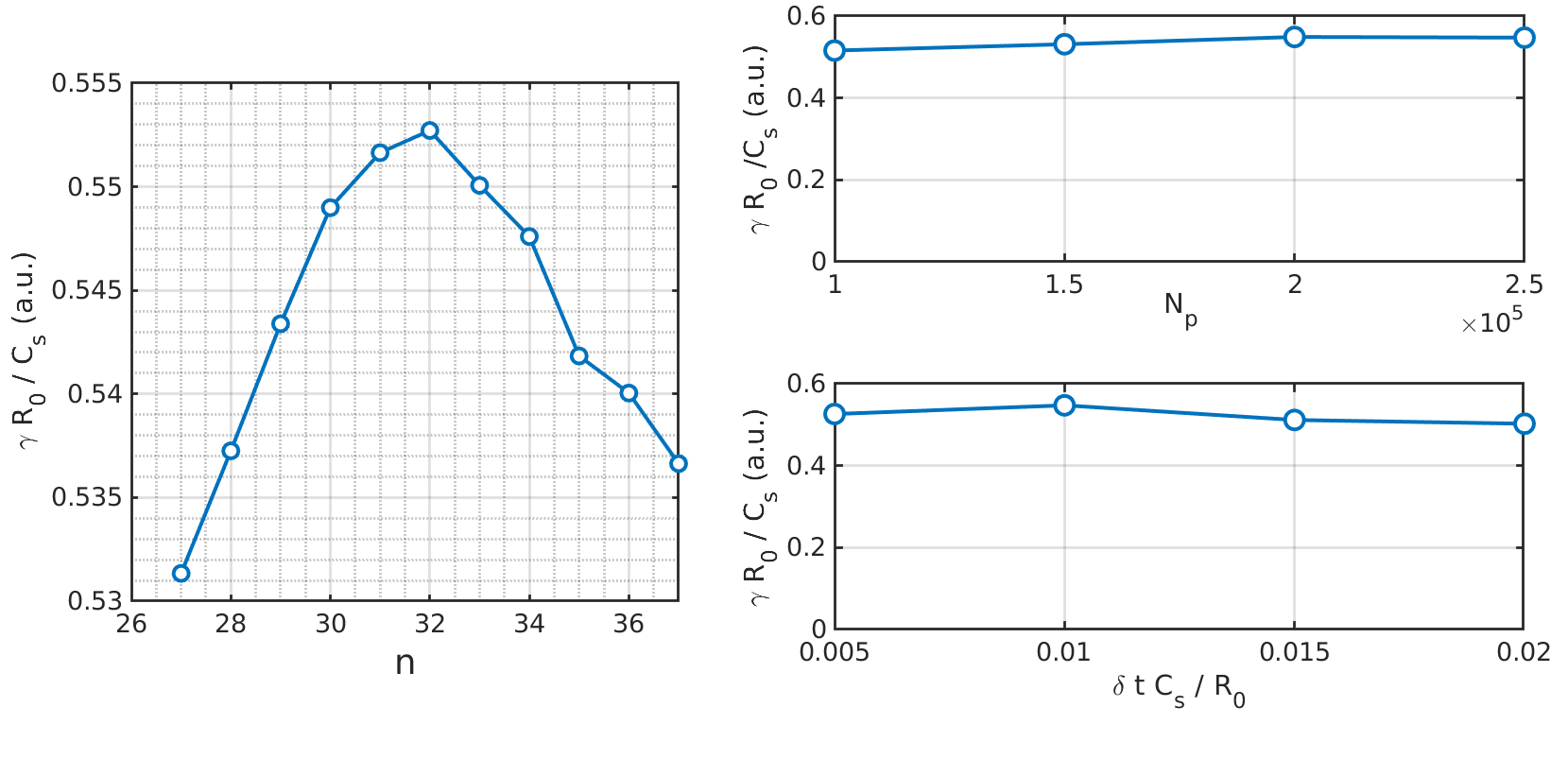} \\
    {(a)} \hspace{6cm} {(b)} }
    \caption{(a) Dispersion relation for the n-modes; (b) Convergence
             of growth rates w.r.t. number of particles per MPI (top)
             and time step (bottom).}
    \label{Fig:Dispersion}
\end{figure}

The convergence of the simulation results in terms of growth rate
concerning the total number of marker particles,
and the time step sizes are given in Fig.~\ref{Fig:Dispersion}(b). 
We find that the growth rate saturates with the ITG amplitude maximum
localizing at the radial location $r\sim 1.19 \; R_0$ and a radial
full-width half maximum of $r \sim 0.01 \; R_0$ (see 
Fig.~\ref{Fig:Growth}(a)). Convergence w.r.t. the marker particle
numbers show that the growth rate saturates for total particles
$\gtrsim 25.6 \times 10^6$ , which is approximately $6$ particles per
triangle. Also, the growth rate stabilizes for the time step 
$\lesssim 0.01 (R_0/C_s)$.

%---------------------------------------------------

\section{Conclusion}
\label{Sec:Conclusions}

We describe the detailed implementation of various modules of G2C3, a gyrokinetic
PIC code to study microinstabilities of fusion plasma confinement in a tokamak system.
The code employs a neural network to perform gather-scatter operations in 
cylindrical coordinates and avoids the problems one encounters by the use of field
line coordinates at the last closed flux surface. We also use a neural network to
locate the triangle from the mesh which encompasses the particle. In G2C3, we
implement a FEM-based gyrokinetic Poisson solver. We benchmark G2C3
by reproducing the linear ITG mode in the core region, with adiabatic electrons, for a Cyclone
test case in the realistic DIII-D geometry.
This result is the first step for G2C3 towards achieving a whole plasma
volume simulation in the future.

%%%%%%%%%%%%%%%%%%%%%%%%%%%%%%%%%%%%%%%%%%%%%%%%%%%%%%%%%%%%%%
\section*{Acknowledgements}
This material is based upon work supported by the U.S.
Department of Energy, Office of Science, Office of Fusion
Energy Sciences, using the DIII-D National Fusion Facility, a DOE
Office of Science user facility, under Award DE-FC02-04ER54698.
This work is supported by National Supercomputing Mission (NSM) (Ref No: DST/NSM/R\&D\_HPC\_Applications/2021/4), Board of Research in Nuclear Sciences (BRNS Sanctioned nos. 39/14/05/2018-BRNS and 57/14/04/2022-BRNS), Science and Engineering Research Board EMEQ program (SERB sanctioned nos. EEQ/2017/000164 and EEQ/2022/000144) and Infosys Young Investigator award. The results presented in this work have been simulated on ANTYA cluster at Institute of Plasma Research, Gujarat, SahasraT and Param Pravega supercomputer at Indian Institute of Science, Bangalore, India.\\

\section*{Disclaimer}
This report was prepared as an account of work sponsored by an agency of the United States Government. Neither the United States Government nor any agency thereof, nor any of their employees, makes any warranty, express or implied, or assumes any legal liability or responsibility for the accuracy, completeness, or usefulness of any information, apparatus, product, or process disclosed, or represents that its use would not infringe privately owned rights. Reference herein to any specific commercial product, process, or service by trade name, trademark, manufacturer, or otherwise does not necessarily constitute or imply its endorsement, recommendation, or favoring by the United States Government or any agency thereof. The views and opinions of authors expressed herein do not necessarily state or reflect those of the United States Government or any agency thereof.
\\

\appendix

\end{document}